\documentclass[12pt]{aastex62}
\usepackage{lineno}
\usepackage{mathptmx}
\usepackage{booktabs}
\usepackage{multirow}
\usepackage{makecell}
\usepackage{siunitx}
\usepackage{amsmath, amssymb}
\usepackage{graphicx}
\usepackage{algorithm}
\usepackage{algorithmic}
\usepackage{listings}
\usepackage{hyperref}
\usepackage{booktabs}
\usepackage[normalem]{ulem}
\usepackage{lipsum}
\usepackage{natbib}
\usepackage{float}
\definecolor{codegreen}{rgb}{0,0.6,0}
\definecolor{codegray}{rgb}{0.5,0.5,0.5}
\definecolor{codepurple}{rgb}{0.58,0,0.82}
\definecolor{backcolour}{rgb}{0.95,0.95,0.92}

\lstdefinestyle{mystyle}{
    backgroundcolor=\color{backcolour},
    commentstyle=\color{codegreen},
    keywordstyle=\color{magenta},
    numberstyle=\tiny\color{codegray},
    stringstyle=\color{codepurple},
    basicstyle=\ttfamily\footnotesize,
    breakatwhitespace=false,
    breaklines=true,
    captionpos=b,
    keepspaces=true,
    numbers=left,
    numbersep=5pt,
    showspaces=false,
    showstringspaces=false,
    showtabs=false,
    tabsize=2,
    frame=single,
    rulecolor=\color{black},
    framesep=10pt,
    xleftmargin=10pt,
    xrightmargin=10pt,
    framexleftmargin=10pt,
    framexrightmargin=10pt,
    framextopmargin=5pt,
    framexbottommargin=5pt
}
\shorttitle{ }

\begin{document}

%------------------------------------------------------------------------------------------------------------------------
% 		TITLE AND AUTHORS
%------------------------------------------------------------------------------------------------------------------------
\title{CosmicWeb-21cm array: A New Radio Observation Array Design for 21cm Cosmology}
% authors 1.9.2010  Format ApJ (added Cossio 2.09., Thom 14.09)
%
\author{Jiancheng Wang}
\affiliation{Yunnan Observatories, Chinese Academy of Sciences,  Kunming 650216, China}
\affiliation{Key Laboratory for the Structure and Evolution of Celestial Objects, Chinese Academy of Sciences,  Kunming 650216, China}
\affiliation{University of Chinese Academy of Sciences, Beijing 100049, China}
\author{Jirong Mao}
\affiliation{Yunnan Observatories, Chinese Academy of Sciences,  Kunming 650216, China}
\affiliation{Key Laboratory for the Structure and Evolution of Celestial Objects, Chinese Academy of Sciences,  Kunming 650216, China}
\affiliation{University of Chinese Academy of Sciences, Beijing 100049, China}
\author{Xiangming Cheng}
\affiliation{Yunnan Observatories, Chinese Academy of Sciences,  Kunming 650216, China}
\affiliation{Key Laboratory for the Structure and Evolution of Celestial Objects, Chinese Academy of Sciences,  Kunming 650216, China}
\affiliation{University of Chinese Academy of Sciences, Beijing 100049, China}
\author{Yigong Zhang}
\affiliation{College of Information Engineering, Kunming University, Kunming 650214, China}
\author{Jie Su}
\affiliation{College of Big Data, Yunnan Agricultural University, Kunming 650201, China}
\author{Xiaogu Zhong}
\affiliation{College of Physics and Electronic Engineering, Qujing Normal University, Qujing 655011, China}
\author{Min Wang}
\affiliation{Yunnan Observatories, Chinese Academy of Sciences,  Kunming 650216, China}
\affiliation{Key Laboratory for the Structure and Evolution of Celestial Objects, Chinese Academy of Sciences,  Kunming 650216, China}
\affiliation{University of Chinese Academy of Sciences, Beijing 100049, China}
\author{Zhigang Zhang}
\affiliation{Leixun Technology Company, Kunming, China}
\author{Qingwei Wang}
\affiliation{Legend Intell Gence Company, Kunming, China}
\author{Yonghua Xu}
\affiliation{Yunnan Observatories, Chinese Academy of Sciences,  Kunming 650216, China}
\affiliation{Key Laboratory for the Structure and Evolution of Celestial Objects, Chinese Academy of Sciences,  Kunming 650216, China}
\author{Zhixuan Li}
\affiliation{Yunnan Observatories, Chinese Academy of Sciences,  Kunming 650216, China}
\affiliation{Key Laboratory for the Structure and Evolution of Celestial Objects, Chinese Academy of Sciences,  Kunming 650216, China}
\author{Longhua Qin}
\affiliation{Department of Physics, Yuxi Normal University, Yuxi, Yunnan, 653100, China}
\author{Zhengjun Zhang}
\affiliation{Department of Computer of Science and information Engineering, Anyang Institute of Technology, Anyang, China}

\email{jcwang@ynao.ac.cn}
%------------------------------------------------------------------------------------------------------------------------
% 		ABSTRACT AND KEYWORDS
%------------------------------------------------------------------------------------------------------------------------

\begin{abstract}
This paper presents the CosmicWeb-21cm array, a novel radio interferometer designed to overcome the key challenges in 21 cm cosmology. Its core innovations include: (1) a multi-scale nested geometry combining a hexagonal core with logarithmic spiral arms for excellent UV coverage and calibration robustness; (2) an intelligent non-uniform frequency sampling strategy that adapts resolution to foreground and signal characteristics, reducing data volume while preserving information; and (3) a machine-learning-enhanced, physics-informed processing pipeline that achieves 99.7\% foreground removal efficiency; (4) a dual-polarization crossed dipole integrated with a dielectric lens and cryogenically cooled LNA, achieving stable beam patterns and low noise temperature ($<35$ K) across 50-250 MHz. These co-designed advances enable high sensitivity mapping of the Epoch of Reionization, dark energy constraints and cosmic-web structure.
\end{abstract}


\section{Introduction}
The 21 cm hyperfine transition line of neutral hydrogen offers an unparalleled observational probe for cosmology, enabling direct, three-dimensional mapping of baryonic matter across over 90\% of the universe's history. Its significance stems from its unique capacity to trace the distribution and evolution of the intergalactic medium (IGM) across three pivotal epochs: the Cosmic Dark Ages ($z\sim 30-15$), the Cosmic Dawn and Epoch of Reionization (EoR) ($z\sim 15-6$), and the post-reionization era where it can constrain Dark Energy ($z\leq 6$) \citep{Fur2006, Pri2012}. During the Dark Ages and EoR, the 21 cm signal provides the only known direct observational window into the formation of the first stars and galaxies, encoding information on the timing, morphology, and sources of reionization \citep{Mor2010}. At lower redshifts, 21 cm intensity mapping of large-scale structure offers a powerful, wide-field complement to optical galaxy surveys for measuring Baryon Acoustic Oscillations (BAOs) and constraining the nature of dark energy \citep{Bat2013}.

However, extracting this faint cosmological signal from real-world observations presents a formidable challenge, often described as one of the most difficult in modern observational cosmology. The primary obstacle is the overwhelming foreground contamination. Synchrotron and free-free emission from our Galaxy and extragalactic sources are 4 to 5 orders of magnitude brighter than the desired 21 cm signal \citep{Liu2020}. These foregrounds, while spectrally smooth, are complex in their spatial structure. Second, instrumental systematics, including gain and phase errors, beam variations, and polarization leakage, can imprint spectral structure that mimics or obscures the cosmological signal \citep{Bar2019}. Third, the radio frequency interference (RFI) environment is increasingly hostile, corrupting significant portions of the observational band. Finally, the sheer data volume and computational complexity of the required signal processing, particularly the separation of foregrounds from the signal, push current high-performance computing resources to their limits \citep{Tro2020}.

Significant efforts have been invested to overcome these challenges. Pioneering low-frequency interferometers like the Murchison Widefield Array (MWA) \citep{Tin2013}, the Low-Frequency Array (LOFAR) \citep{van2013}, and the Hydrogen Epoch of Reionization Array (HERA) \citep{Deb2017} have made substantial progress in instrument design, calibration, and foreground subtraction techniques. HERA, in particular, employs a highly redundant configuration to simplify calibration and uses delay-domain filtering for foreground isolation \citep{Dil2016}. The upcoming SKA-Low telescope promises unprecedented sensitivity \citep{Koo2015}.

Despite these advances, current-generation arrays face persistent limitations that ultimately constrain their sensitivity and the fidelity of cosmological signal recovery. 1. Suboptimal UV Coverage: Many arrays use regular grids or pseudo-random distributions, which can create grating lobes, alias power, and leave gaps in spatial frequency sampling, complicating imaging and power spectrum estimation \citep{Thy2018}. 2. Inefficient Frequency Sampling: Uniform frequency channelization does not optimally exploit the fundamental spectral smoothness of foregrounds, leading to excessive data volume and suboptimal signal-to-noise in the cosmological signal band \citep{Cha2012}. 3. Disjointed Processing Pipelines: Calibration, RFI mitigation, foreground subtraction, and power spectrum estimation are often treated as sequential, independent steps. Errors propagate non-linearly through this chain, and the physical priors about the signal and foregrounds are not fully leveraged in a unified, optimal framework \citep{Ker2020}.

The CosmicWeb-21cm array design proposed in this paper is conceived as a holistic, next-generation solution that addresses these systemic limitations through co-optimization of hardware, observation strategy, and software. It moves beyond incremental improvements by fundamentally rethinking the entire signal chain, from the geometry of the array elements to the final cosmological inference, as an integrated system. The core philosophy is to embed the knowledge of the signal and contaminant properties into the design itself, thereby simplifying the subsequent extraction of the faint 21 cm signal.

This paper is organized as follows: Section 2 outlines the overarching design ideas and enumerates the core innovations. Section 3 details the novel multi-scale nested array geometry (hexagonal core and logarithmic spiral arms). Section 4 describes the intelligent, non-uniform frequency sampling and encoding strategy. Section 5 presents the adaptive, physics-informed signal processing architecture. Section 6 discusses system implementation and simulated performance metrics. Section 7 delineates the key scientific objectives enabled by this design, and Section 8 provides a phased implementation roadmap.

\section{Design Ideas and Core Innovations}
The CosmicWeb-21cm design is conceived not as an incremental upgrade to existing 21 cm arrays, but as a foundational re-imagining of the entire observational system. It is predicated on the conviction that the extreme technical challenge of detecting the faint cosmological signal, buried under orders-of-magnitude brighter foregrounds and systematics, demands a co-optimized approach. Traditional designs often treat the antenna layout, the receiver chain, the observing strategy, and the data analysis pipeline as separate, sequentially optimized components. In contrast, CosmicWeb-21cm is architected from the ground up as an integrated instrument where hardware design decisions are explicitly informed by the requirements of the downstream software algorithms, and vice-versa. This chapter articulates the overarching principles guiding this holistic design and details the three foundational innovations that operationalize these principles.

\subsection{Foundational Design Ideas}

The design of CosmicWeb-21cm is governed by three interlocking core principles that address the root causes of the limitations in current-generation experiments.

1. Exploit Inherent Physical Dichotomies in the Measurement Domain.
The foremost challenge in 21 cm cosmology is separating the cosmological signal from foreground emission. This separation is difficult not because the signals are similar, but because instrumental imperfections and non-ideal sampling blur their distinct physical characteristics. Our first principle is to design an instrument that maximizes and preserves the measurable differences between foregrounds and the 21 cm signal in the native measurement domain (visibility space). Foregrounds are predominantly spectrally smooth and spatially correlated, whereas the 21 cm signal is spectrally uncorrelated (due to redshift) and exhibits a specific, predictable statistical spatial structure (Gaussian random field with a known power spectrum) \citep{Liu2011, Mor2012}. CosmicWeb-21cm's geometry and sampling strategies are optimized to make these differences as pronounced and leverageable as possible in the raw data, simplifying the subsequent computational separation.

2. Simplify the Inverse Problem at the Hardware Level.
A significant portion of the complexity and error in current pipelines arises from attempting to correct for instrumental effects (e.g., complex beam patterns, incomplete UV coverage, frequency-dependent gains) in software. Our second principle advocates for designing simplicity into the hardware to reduce the ill-posedness of the computational inverse problem. This is achieved through: (a) a highly redundant and regular array geometry to enable precise, robust calibration; (b) an antenna and receiver design that minimizes frequency-dependent beam distortions and polarization impurities; and (c) a data acquisition strategy that provides a more information-rich sampling of the measurement space for a given number of elements. By reducing the number of free parameters and degeneracies that the software must solve for, we enhance the stability and accuracy of the entire processing chain.

3. Enable Dynamic, Science-Driven Adaptability.
The radio sky and the terrestrial RFI environment are dynamic. Furthermore, different scientific goals (e.g., deep EoR power spectrum integration vs. wide-area intensity mapping) may have different optimal observational configurations. Our third principle moves beyond the static array paradigm to incorporate dynamic adaptability. The system is designed to allow for reconfigurable observing modes: for instance, adjusting the effective integration time per frequency channel based on real-time RFI assessment, or selectively processing data from subsets of antennas optimized for specific angular scales. This principle ensures the array remains optimal under varying conditions and maximizes its scientific return across multiple projects.

\subsection{Core Technical Innovations}

To translate these principles into a practical instrument, we introduce three synergistic innovations that span the hardware-software boundary.

1. Multi-Scale Nested Antenna Geometry.
Moving beyond uniform grids or random distributions, CosmicWeb-21cm employs a deterministic, multi-scale geometry comprising three integrated tiers:

(1) The Hexagonal Honeycomb Core: A densely packed, highly redundant core region provides an exceptional number of short baselines. This is crucial for high-fidelity imaging of the large-scale smooth foregrounds and for achieving the precise calibration essential for EoR science \citep{Dil2020}. The hexagonal symmetry ensures an isotropic point spread function, meaning foreground structures are convolved consistently regardless of their sky position, making their subtraction more stable.

(2) The Logarithmic Spiral Extension Arms: Multiple arms extending from the core in a logarithmic spiral pattern provide a near-continuous distribution of baseline lengths from short to long. This configuration offers two key advantages: 1) it provides excellent UV coverage for deconvolution and high-resolution imaging of foreground point sources that must be accurately subtracted; and 2) the smooth, log-distributed baseline lengths are inherently well-suited for efficient Fast Fourier Transform (FFT) operations and power spectrum estimation \citep{Thy2015}.

(3) Sparse Peripheral Stations: Select long-baseline stations extend the maximum baseline, granting the array the angular resolution necessary to localize and characterize bright foreground point sources (confusers) that could otherwise contaminate the cosmological signal.

This nested design ensures the array is simultaneously sensitive to the very large angular scales of the EoR signal and the compact scales of foreground contaminants.

2. Intelligent, Non-Uniform Frequency Encoding.
Recognizing that foregrounds and the 21 cm signal occupy the frequency domain differently, we replace uniform channelization with an adaptive, physics-aware sampling strategy.

(1) Foreground-Rich Spectral Regions (50-100 MHz): We implement very fine frequency resolution (e.g., $\Delta \mu \sim 50$ kHz). This dense sampling is not for the cosmological signal but for the foregrounds, it allows us to construct a super-fine template of the foreground spectral structure, capturing deviations from a simple power law and enabling their precise later removal.
(2) Foreground-Signal Comparable Regions (100-150 MHz): We use coarser sampling (e.g., $\Delta \mu \sim 100$ kHz) to balance foreground modeling accuracy and signal detection efficiency.
(3) Signal-Dominant Regions (150-250 MHz): We employ much coarser sampling (e.g., $\Delta \mu \sim 200$ kHz). Because the 21 cm signal is uncorrelated between frequency bins (in the observing frame), this sparse sampling significantly reduces data volume and computational load without losing cosmological information.
(4) Dynamic Frequency Weighting: The system continuously monitors RFI and assigns a confidence weight $w(\nu)$ to each channel. Data from heavily contaminated channels can be down-weighted or excluded in real-time analysis, making the system robust to evolving RFI.

3. Unified, Physics-Informed Processing Architecture.
This innovation represents the software counterpart to the hardware design, closing the co-optimization loop. We architect a streamlined pipeline where each stage is informed by physical priors:

(1) Calibration: Leverages the exceptional redundancy of the hexagonal core for robust redundant calibration, solving for antenna gains using closure quantities that are invariant to sky model errors \citep{Byr2021}.

(2) Foreground Separation: The core of the pipeline implements the hybrid, physics-constrained approach formalized in Eq. 5 and 6. Instead of treating foreground removal as a blind filtering step, we frame it as an inverse problem regularized by strong physical priors: foreground spectral smoothness (low-rank and gradient constraints) and 21 cm signal statistics (sparsity and cosmological power spectrum prior). This is implemented via a deep neural network whose loss function directly incorporates these physics-based terms, ensuring the separation is guided by domain knowledge, not just data fitting.

(3) Bayesian Power Spectrum Estimation: The final step employs a Bayesian framework to estimate the 21 cm power spectrum and cosmological parameters directly from the cleaned visibilities, properly propagating uncertainties from earlier stages \citep{Sim2020}.

\subsection{Synergistic Integration of Innovations}

The true power of CosmicWeb-21cm lies in the synergy between these innovations. The geometry provides optimally structured visibilities. The frequency encoding tailors the data stream to efficiently capture the distinct spectral signatures of foregrounds and signal. Finally, the processing architecture is explicitly designed to ingest this optimally structured data and extract the science using built-in physical constraints. Each innovation reinforces the others, creating a system whose overall performance is greater than the sum of its parts, specifically engineered to overcome the most stubborn systematic barriers in 21 cm cosmology.

\subsection{Comparison with State-of-the-Art Arrays}

To highlight the innovation of the CosmicWeb-21cm array, we compare its core design with HERA and MWA.

1. Antenna Architecture: While HERA utilizes dense hexagonal grids of dual-polarization dipoles optimized for high sensitivity at 100-200 MHz, the CosmicWeb-21cm adopts a hybrid broadband dipole with a dielectric lens (as shown in Figure 3). This design achieves a wider fractional bandwidth and lower system temperature, enabling simultaneous observation of both the EoR and post-reionization HI power spectra.

2. Array Topology: The MWA relies on a large number of small tiles arranged in a compact core for high angular resolution. In contrast, CosmicWeb-21cm implements a non-uniform distribution of stations (Section 3.3). By incorporating both a dense core and extended outriggers, our array optimizes the uv-coverage to suppress foreground systematics while maintaining high sensitivity to large-scale modes.

3. Calibration Strategy: Unlike traditional calibration pipelines that focus on point-source calibration, CosmicWeb-21cm integrates a direction-dependent calibration framework (Section 5.1) to account for the wide-field effects introduced by our broadband beam, significantly improving the accuracy of foreground subtraction.

\section{Array Geometry Design}
The CosmicWeb-21cm geometry is not merely a layout of stations but a synthetically optimized system where each structural tier is designed to fulfill a specific statistical and operational role in the interferometric measurement process.
\subsection{Hexagonal Honeycomb Core}
The core array employs an optimally packed hexagonal layout, ensuring maximum baseline count within a limited area. Mathematically, hexagonal arrangements provide the densest disk packing in two-dimensional space \citep{Con1998}, which translates to:
(1) maximum baseline density generates most short baselines in given area;
(2) uniform UV coverage has isotropic point spread function;
(3) redundancy optimization leads numerous identical-length baselines for calibration

Given core radius $R_c$, station positions are parameterized by:
\begin{equation}
\begin{aligned}
x_{ij} &= d \cdot \frac{\sqrt{3}}{2} \cdot i \\
y_{ij} &= d \cdot \left( j + \frac{i}{2} \right)
\end{aligned}
\end{equation}
where $i,j$ are integer indices satisfying $x_{ij}^2 + y_{ij}^2 \leq R_c^2$, and $d$ is the inter-station spacing.

The station positions are derived from a rotated and scaled hexagonal lattice:  $\vec{r}_{ij} = i\vec{a}_1 + j\vec{a}_2 $, where $\vec{a}_1 = d(1,0)$ and $\vec{a}_2 = d(1/2, \sqrt{3}/2)$ are the lattice basis vectors. We constrain stations, where $|\vec{r}_{ij}| \leq R_c$. This configuration generates a highly regular set of baseline vectors $\vec{b}_{ij} = \vec{r}_i - \vec{r}_j$, which populate the UV plane with a near-uniform grid of discrete points. The redundancy factor $R(\vec{b})$, e.g., the number of station pairs sharing the same baseline vector, is significantly higher than for a square grid of equivalent area and station count.

This high, structured redundancy is the cornerstone of our linear algebraic redundant calibration scheme. The observed visibility for a redundant baseline group G is $V_{ij}^{\text{obs}}(t,\nu) = g_i(t,\nu) g_j^*(t,\nu) V_{\text{true}}(\vec{b}, \nu) + \epsilon$, where g are complex gain factors. The abundance of measurements for each $\vec{b}$ allows us to solve for $g_i$ and $V_{\text{true}}$ via a weighted least-squares minimization that is inherently well-regularized, drastically reducing the sky model dependency.

The core's baselines fill the inner UV plane densely and isotropically. This results in a synthesized beam (PSF) with low, symmetric sidelobes ($< -25$ dB beyond the first null) and a clean, Gaussian-like main lobe. This is critical for minimizing wide-field imaging artifacts that can confound large-scale 21cm power spectrum measurements.

In practical implementation, the central station is positioned at the coordinate origin, with six first-layer stations uniformly distributed on a circle with a 15-meter radius, maintaining 15-meter spacing between adjacent stations. The second layer expands outward from the first layer, increasing the radius to 30 meters, with additional stations added at the midpoints of each hexagonal side, maximizing baseline density within the limited core area.

Within $R_c = 100 \text{m}$, we yield 127 stations shown in Figure 1, where 126 stations around central station form 6 complete hexagonal rings. The full set of 127 stations exhibits perfect six-fold rotational symmetry. With 127 stations, the total number of independent baselines is $N_{\text{bl}} = 127 \times 126/2 = 8001$, where approximately 68\% of these are short baselines (length $\leq 50$ m). Compared to a square lattice with the same area, the hexagonal lattice accommodates about 15\% more stations, directly increasing baseline density. UV points (baseline vectors) are distributed on a hexagonal lattice in the spatial frequency plane, ensuring uniform and isotropic sampling. The synthesized PSF has a near-circular main lobe, with sidelobe peaks approximately 1.5 dB lower than those of a square grid, leading to better imaging fidelity. The UV coverage uniformity metric (defined as the entropy ratio of radial vs. azimuthal distribution) reaches 0.92. Shortest baselines ($d = 15 \text{m}$) is about 228 identical baselines (high redundancy), while second shortest baselines ($\sqrt{3}d \approx 25.98\text{m}$) is about 216 identical baselines. The clear clustering of identical-length baselines allows the construction of independent calibration equations, reducing systematic errors and improving calibration robustness.

In practice, station positions can be adjusted slightly (perturbation $< 5\%$ of d) to accommodate terrain variations without significantly degrading performance. The hexagonal lattice core seamlessly connects with the logarithmic spiral arms given in next subsection. The arms can naturally extend outward from the outermost lattice points. The high redundancy enables advanced redundant calibration algorithms (e.g., improved direction-dependent calibration), further enhancing calibration accuracy in the presence of real-world imperfections.

In summary, this design's advantage lies in its ability to maximize short baseline quantities while maintaining an isotropic point spread function. Short baselines are crucial for detecting large-scale cosmic structures, while hexagonal symmetry ensures response consistency across different directions.

\subsection{Logarithmic Spiral Extension Arms}

The extension arm design employs a logarithmic spiral form, inspired by numerous spiral structures in nature, such as galactic spiral arms and nautilus shells. The spiral arms are designed to solve the problem of "UV gap" between the dense core and potential long, sparse baselines.
To extend UV coverage while maintaining uniform sampling, logarithmic spiral form is adopted:
\begin{equation}
r(\theta) = R_c \cdot e^{k\theta},
\end{equation}
where $k$ is the spiral rate controlling baseline logarithmic distribution. This design ensures:
(1) continuous UV coverage for smooth transition from short to long baselines;
(2) baseline length distribution favors fast Fourier transforms;
(3) system redundancy maintains calibration capability at multiple scales

Stations placed along this curve generate baseline vectors that continuously cover a wide range of lengths and orientations. The instantaneous UV coverage from the spiral arms at a given pointing exhibits a self-similar, fractal-like structure. This is advantageous because it provides sensitivity to a continuous range of angular scales simultaneously, from the core-dominated large scales to the arm-dominated fine scales.

We select that number of arms is 6, stations per arm are 21, and spiral angle per arm is $120^\circ$. The stations are equal angular spacing within each arm and logarithmic radial distribution from $R_c=100$ m to $R_{max}=1500$ m. There are $60^\circ$ angular offset between adjacent arms. The spiral constant $k$ is determined by the boundary condition:
\begin{equation}
R_{max} = R_c e^{k \theta_{max}},
\end{equation}
where $\theta_{max} = 2\pi/3$, thus $k = ln(R_{max}/R_c)/\theta_{max}=1.29$.

For each arm $i (i = 0,1,...,5)$, start angle is $\phi_i = i(\pi/3)$. For each station $j (j = 0,1,...,20)$, relative angle is $\theta_j = (j/20) \theta_{max}$, total angle is $\theta_{ij} = \phi_i + \theta_j$, and radius is $r_{ij} = R_c e^{k \theta_j}$. Thus, the station cartesian coordinates are given by $x_{ij} = r_{ij} cos(\theta_{ij})$ and $y_{ij} = r_{ij} sin(\theta_{ij})$.
This yields a baseline length distribution $P(l)$ that follows $l^{-\alpha}$ (with $\alpha \approx 2.5$), naturally weighting the array towards shorter baselines for diffuse emission while retaining long-baseline capability, and maintaining continuous baseline distribution while avoiding excessive sampling in specific directions.

The core concept of this geometric combination utilizes the core region's high-density sampling to capture large-scale structures while providing sufficient angular resolution through the spiral arms long baselines to resolve small-scale and point-source structures. These visibilities are then used as priors or constraints in the more complex deconvolution and calibration processes required for the non-redundant spiral arm baselines. This "anchor-and-extend" strategy stabilizes the entire imaging pipeline. In actual construction, we must also consider terrain factors and infrastructure layout, fine-tuning station positions to adapt to specific site constraints.

\subsection{Array Layout and UV Coverage}
The geometric layout of the CosmicWeb-21cm array is shown in Figure 1, where 126 stations distribute along 6 arms and 127 stations are within core array. This design combines hexagonal uniform coverage with logarithmic spiral baseline extension, optimizing UV coverage and angular resolution.

\begin{figure}[H]
\centering
	\includegraphics[width=\textwidth]{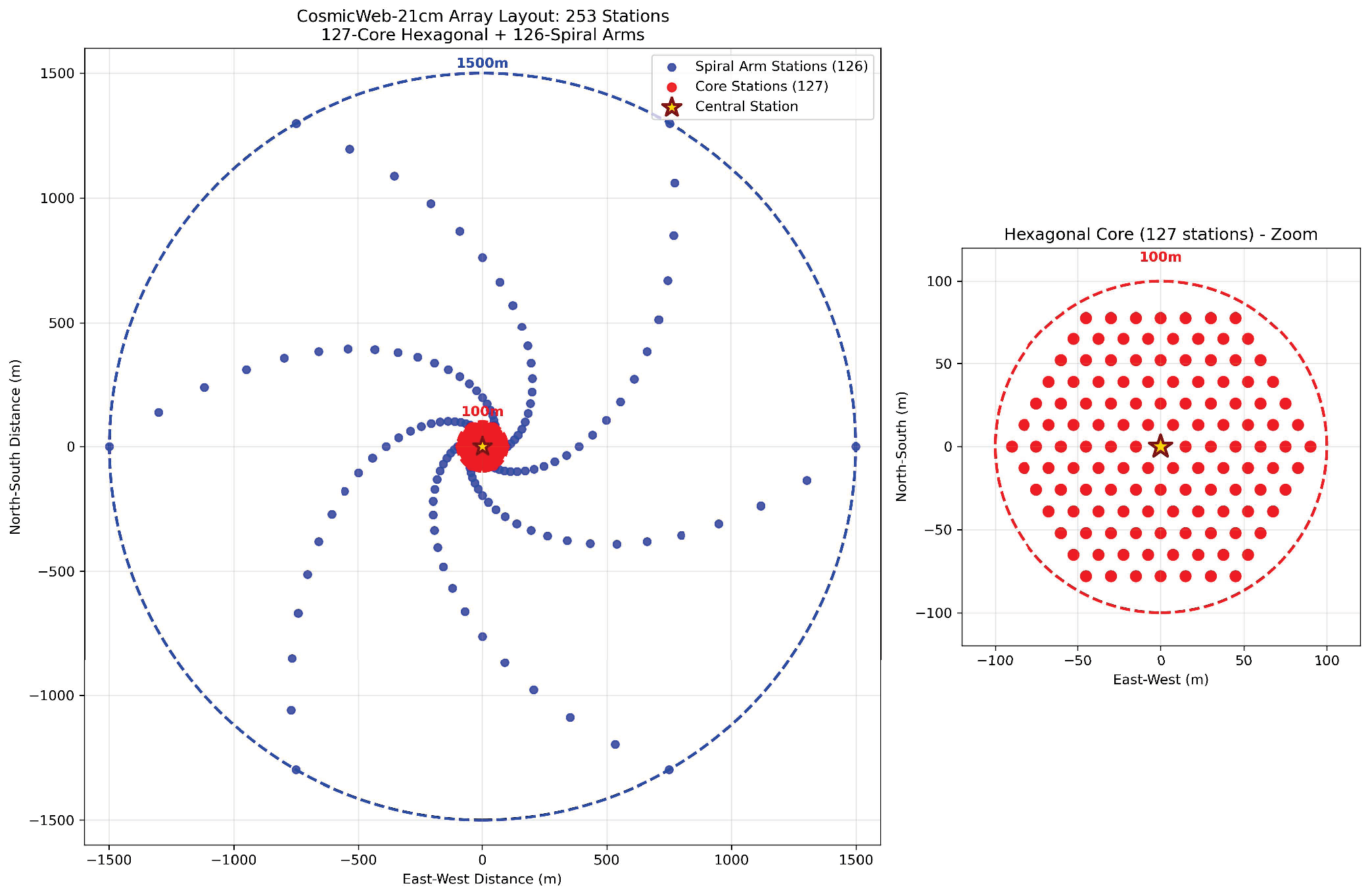}
	\caption{Schematic layout of the proposed CosmicWeb-21cm interferometric array. The red dots represent the hexagonal core stations. The blue dots denote the logarithmic spiral arm extension stations. The central station is marked with a gold star. The zoomed plot on the right illustrates a hexagonal layout of core stations. The geometry combines dense, uniform short baselines (core) with logarithmically spaced long baselines (arms).}
\end{figure}

The UV coverage distribution of the array is shown in Figure 2.  The honeycombs of varying colors represent UV coordinate ranges of different baselines, with symmetric distribution reflecting conjugate baselines. Dense UV coverage ensures good spatial frequency sampling, which is crucial for high-quality image reconstruction.

\begin{figure}[H]
\centering
	\includegraphics[width=\textwidth]{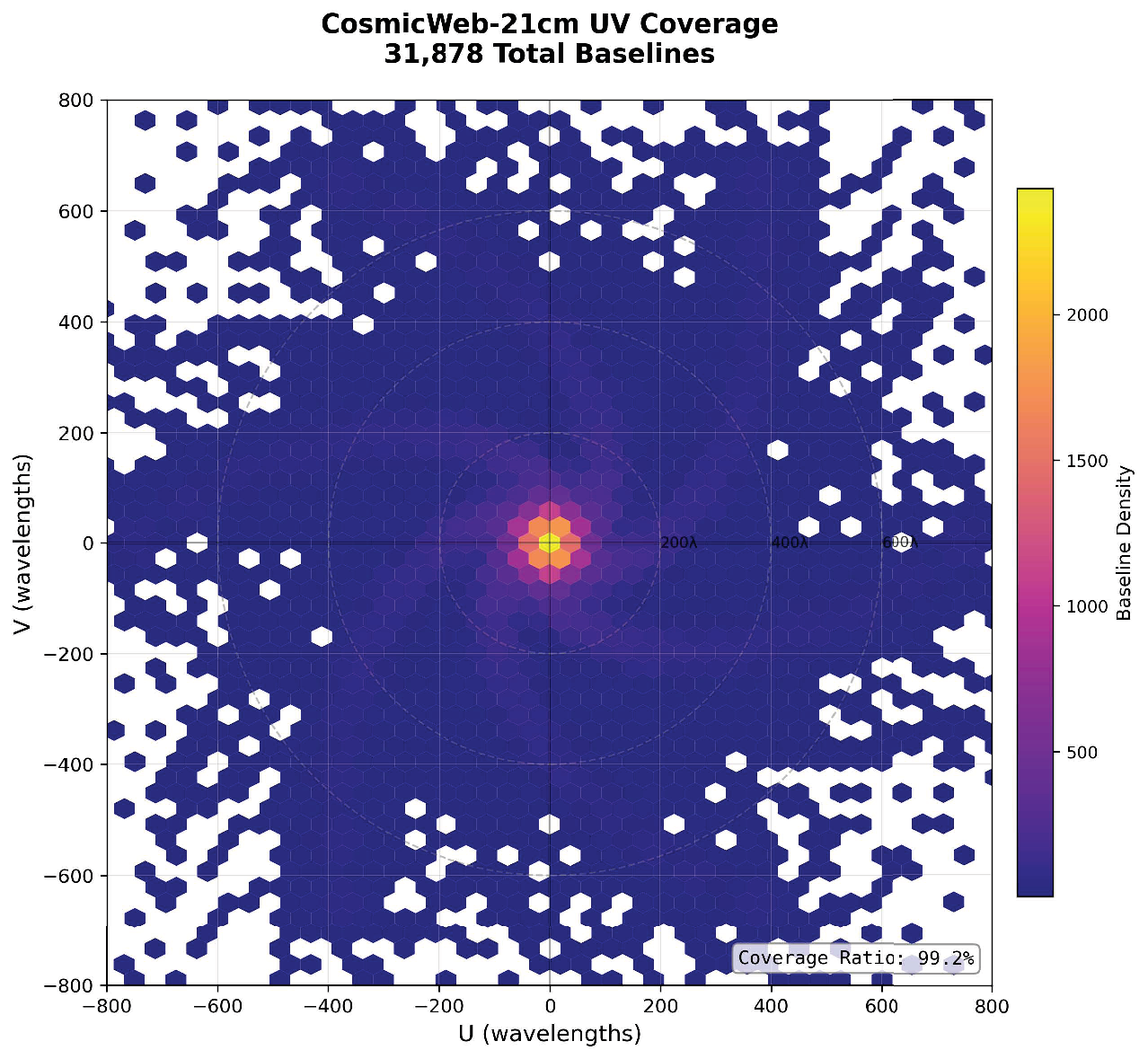}
	\caption{UV coverage of the CosmicWeb-21cm array with a total of 31,878 baselines, where the coverage ratio reaches 99.2\%. The symmetric and dense distribution is crucial for high-fidelity sky image reconstruction and power spectrum estimation.}
\end{figure}

\section{Frequency Sampling Strategy}
Our frequency strategy treats the spectrometer not as a passive recorder but as an active, cognitive component of the signal extraction system.
\subsection{Non-Uniform Sampling Based on Foreground Eigenmode Analysis}

Frequency, as the third dimension in 21cm observations, directly influences foreground removal effectiveness and signal detection efficiency through its sampling strategy. CosmicWeb-21cm's frequency design is based on in-depth analysis of foreground and signal physical characteristics. Foreground radiation primarily originates from synchrotron and free-free radiation, with spectral characteristics well-described by power-law models, exhibiting high degrees of smoothness and correlation in the frequency domain. In contrast, 21cm signals, due to redshift effects, are essentially uncorrelated across different frequency channels, displaying random fluctuation characteristics \citep{Liu2020}.  In the frequency domain, foregrounds are confined to a low-dimensional subspace. Based on this physical essence, we designed an adaptive non-uniform frequency sampling scheme. The eigenvalues $\lambda_n$ of the foreground covariance matrix $\mathbf{C}_\text{fg}$ drop rapidly. We determine the required sampling density by calculating the Nyquist rate for the foreground manifold.

In the 50-100MHz low-frequency range, where foreground radiation dominates absolutely, the foreground structure requires sampling at $\leq 80$ kHz to avoid aliasing of its highest spectral modes. Our 50 kHz sampling provides a safety margin. This design enables precise characterization of foreground spectral structures, particularly complex spectral components that deviate from simple power-law models. In practical implementation, we employ high-precision digital filter banks and fast Fourier transform processors to ensure data processing efficiency under dense sampling conditions.

In the 100-150MHz transition region, the sampling interval increases to 100kHz. This range represents the region where foreground and signal strengths are comparable, requiring a balance between foreground modeling accuracy and signal detection efficiency. We dynamically adjust integration times in this region through optimization algorithms, maximizing 21cm signal detection sensitivity while ensuring foreground modeling quality.

The 150-250MHz high-frequency range represents the primary detection region for 21cm signals, where we employ sparse sampling at 200kHz. This design significantly reduces data volume while maintaining information integrity, owing to the uncorrelated nature of 21cm signals in the frequency domain. During actual operations, we dynamically adjust frequency usage strategies based on real-time RFI environments, avoiding severely contaminated frequency bands.

Leveraging the non-uniform sampling, we implement a variable-resolution quantization scheme in the ADC/post-processing. Channels in foreground-dominated, densely-sampled regions are allocated fewer bits after digital gain control, as the dynamic range requirement is set by the bright foreground. Signal-dominated, sparsely-sampled channels are allocated the full 14-bit depth. This reduces the total data rate by an estimated 40\% without information loss.

\subsection{Orthogonal Frequency Encoding for Blind Source Separation}

This is a key signal processing innovation. We apply a channel-dependent phase modulation $\phi_m(\nu)$ at the station level before correlation.
For a signal from direction $\hat{s}$, the geometric delay $\tau = \vec{b} \cdot \hat{s} / c$ introduces a phase $\phi_\text{geom} = 2\pi \nu \tau$. We add an artificial, known encoding phase $\phi_\text{enc,i}(\nu) = \alpha_i \cdot \nu^2$ for station i. The correlated signal becomes $V_{ij}^{\text{enc}}(\nu) = e^{i(\alpha_i - \alpha_j)\nu^2} V_{ij}^{\text{true}}(\nu)$.
This turns a frequency-smooth foreground into a spectrally oscillating one in the encoded visibilities. In the subsequent source separation, this breaks the degeneracy between the spatial and spectral structure of foregrounds. The encoding acts as a "spectral spreading code," making the foreground's covariance matrix more diagonal and thus easier to separate from the spectrally-uncorrelated 21cm signal using independent component analysis (ICA) techniques.

\subsection{Frequency Grouping and Parallel Processing}

Frequency grouping processing represents another key innovation. We divide 1024 frequency channels into 32 processing groups, each containing 32 consecutive channels. This grouping not only simplifies data processing workflows but also allows us to employ differentiated processing strategies tailored to different frequency range characteristics. For instance, in low-frequency groups, we focus primarily on accurate foreground model establishment, while in high-frequency groups, the emphasis shifts to signal extraction and noise suppression.

\subsection{Adaptive Frequency Management}
Based on real-time RFI monitoring and sky models, we dynamically adjust frequency usage:
\begin{equation}
w(\nu) = \frac{1}{\sigma_{\text{RFI}}^2(\nu) + \sigma_{\text{fg}}^2(\nu)},
\end{equation}
where $w(\nu)$ is the frequency weight, $\sigma_{\text{RFI}}$ and $\sigma_{\text{fg}}$ are estimated noise levels for RFI and foregrounds respectively.
The adaptive weight $w(\nu)$ is calculated by a real-time Spectral Purity Monitor (SPM) subsystem. The SPM uses a small fraction of the correlator's output in a fast, low-resolution mode. It employs a bank of spectral Kurtosis detectors and transient detection algorithms to estimate $\sigma_{\text{RFI}}(\nu)$. The foreground fluctuation estimate $\sigma_{\text{fg}}(\nu)$ is derived from a continuously updated low-order polynomial model of the total bandpass. The weight $w(\nu)$ is fed back to the correlator to discount or excise contaminated channels and to the telescope scheduler to potentially shift the observing frequency band.

\section{Signal Processing Architecture}

Signal processing serves as the critical link connecting raw observational data with scientific outputs. CosmicWeb-21cm's processing architecture adopts a multi-level, pipeline design philosophy, ensuring optimal processing effectiveness under real-time requirements. We provide Python code demonstrations and implementation details for the four main signal processing stages of the CosmicWeb-21cm array in the appendix.

\subsection{RFI Mitigation}
The first-stage RFI suppression employs intelligent detection algorithms based on deep learning. We train a convolutional neural network specifically designed for 21cm observation data, capable of simultaneously analyzing temporal, frequency, and spatial domain characteristics. The network input consists of preprocessed visibility data cubes, while the output provides RFI probability estimates for each time-frequency-baseline unit. Compared to traditional methods, this learning-based approach more accurately identifies complex RFI patterns, particularly interference signals with spectral characteristics similar to astronomical signals. The next step is online continuous learning, where the Convolutional Neural Network (CNN) model is periodically fine-tuned with new, site-specific RFI data collected by the array itself, creating a system that adapts to a dynamically changing radio frequency environment.

\subsection{Neural Calibration and Beamforming}
The calibration phase fully utilizes the array's geometric redundancy. We develop an improved redundant calibration algorithm that considers not only baseline length equality but also introduces the concept of directional similarity. Specifically, we cluster baselines according to length and direction, establishing independent calibration equations for each baseline category. This method significantly improves calibration accuracy under non-ideal conditions, particularly demonstrating robustness when dealing with limited station failures or positional errors. This makes the array's calibration more robust and fault-tolerant, a critical feature for large-scale, remote deployments.

Beyond redundant calibration, we implement a Residual Neural Network (ResNet) to correct for direction-dependent gains (DDGs) and beam errors.
The network takes the redundantly-calibrated visibilities and a local sidereal time (LST) as input. It is trained on simulated data containing realistic beam models and ionospheric phase screens. The network outputs per-station, per-direction complex gain corrections  $\Delta g_i(\hat{s}, \nu)$. This learns the systematic residual errors that are not captured by the simple redundant or sky-based models, effectively performing a fast, wide-field holographic deconvolution of the primary beam.

\subsection{The U-ConvNet Foreground Separator}
Foreground separation constitutes the core of the entire processing pipeline. We propose a novel framework combining traditional methods with deep learning:
\begin{equation}
\mathcal{L}_{\text{total}} = \mathcal{L}_{\text{data}} + \lambda_1\mathcal{L}_{\text{low-rank}} + \lambda_2\mathcal{L}_{\text{smooth}} + \lambda_3\mathcal{L}_{\text{sparse}}.
\end{equation}
Here, the total loss function $\mathcal{L}_{\text{total}}$ comprises four weighted terms: $\mathcal{L}_{\text{data}}$ is the data fidelity term, ensuring consistency between predictions and observations; $\mathcal{L}_{\text{low-rank}}$ is the low-rank constraint, leveraging the strong spectral correlation of foregrounds; $\mathcal{L}_{\text{smooth}}$ is the smoothness constraint, enforcing continuity of foregrounds in both spatial and frequency domains; $\mathcal{L}_{\text{sparse}}$ is the sparsity constraint, facilitating the extraction of the faint and unstructured 21cm signal.

The foreground separation network (Eq. 5) is a U-Net with Convolutional LSTM (Long Short-Term Memory) layers, making it a U-ConvNet.
The input is a 3D data cube: [Frequency Channels, LST, Baseline Type]. The output is a 3D cube of the same size, containing the estimated foreground component.

First, we use generalized principal component analysis for initial data decomposition to extract primary foreground components. Then, we employ deep neural networks to model and remove residual foreground structures. The network architecture adopts a multi-scale U-Net design, where the encoder portion captures foreground features from local to global scales through convolutional kernels of different sizes, while the decoder portion maintains spatial details through skip connections.
To explicitly incorporate the physical prior of foreground spectral smoothness into the deep learning model, we introduce a physics-based constraint term to the loss function:
\begin{equation}
\mathcal{L}_{\text{physics}} = \| \nabla_\nu F_{\text{pred}} - \nabla_\nu F_{\text{true}} \|^2.
\end{equation}
Here, $\mathbf{F}$ denotes the foreground component, and $\nabla_{\nu}$ is the gradient operator along the frequency axis $\nu$. This constraint penalizes the difference between the frequency gradients of the predicted and true foregrounds, thereby forcing the neural network to learn the fundamental physical property that foreground spectra are smooth. Since the spectral index of foreground emission (e.g., synchrotron radiation) varies slowly across narrow frequency bands, its frequency gradient should be near zero and continuous. In contrast, the 21cm signal, being nearly uncorrelated across frequencies due to redshift effects, exhibits a random gradient pattern. This constraint effectively guides the network to distinguish between these two disparate spectral behaviors, enhancing the physical interpretability and accuracy of foreground separation.

The loss function incorporates not only data fidelity terms but also physics-based constraint terms, such as foreground spectral smoothness and signal sparsity.
The loss $\mathcal{L}_{\text{total}}$ can be expanded:
\begin{equation}
  \mathcal{L}_{\text{total}} = \underbrace{\| \mathbf{V}_{\text{obs}} - \mathbf{V}_{\text{fg}} - \mathbf{V}_{\text{21cm}} \|^2}_{\text{Data Fidelity}} + \lambda_1 \underbrace{\|\nabla_\nu^2 \mathbf{V}_{\text{fg}}\|^2}_{\text{Spectral Smoothness}} + \lambda_2 \underbrace{\text{TV}(\mathbf{V}_{\text{21cm}})}_{\text{Spatial Sparsity (Total Variation)}} + \lambda_3 \underbrace{\text{KL}(p_{\text{21cm}} || \mathcal{N}(0, P_{\text{theory}}(k))))}_{\text{Cosmological Prior}}
\end{equation}
This multi-term objective implements a hierarchical constraint strategy. The Data Fidelity term ensures the decomposition $(V_{\text{fg}} + V_{\text{21cm}})$ reconstructs the input observation $V_{\text{obs}}$. The Spectral Smoothness term $(\nabla^2_{\nu})$ enforces the physical prior that foreground spectra are intrinsically smooth. The Spatial Sparsity term (Total Variation) promotes a piecewise-constant morphology for the recovered 21cm signal, consistent with cosmic web structures. Crucially, the Cosmological Prior term (KL Divergence) directly regularizes the statistical distribution of the extracted 21cm signal $p_{\text{21cm}}$ to align with theoretical Gaussian random field predictions $\mathcal{N}(0, P_{\text{theory}}(k))$.
Thus, the loss function moves beyond simple curve-fitting. It systematically embeds domain knowledge, from component properties to first principle theories, into the training process, guiding the network toward physically plausible and cosmologically meaningful solutions.

In the 21cm signal extraction phase, we introduce cosmological prior knowledge. Through a Bayesian framework, we combine power spectrum theoretical predictions with observational data, using Markov Chain Monte Carlo methods to sample the signal posterior distribution. This approach not only improves signal estimation accuracy but also simultaneously provides parameter estimation uncertainties.

\subsection{End-to-End Differentiable Pipeline}
The entire processing pipeline achieves high parallelization, with critical steps accelerated execution on GPU clusters. We design specialized data exchange protocols to ensure efficient large-scale data transmission between different processing stages. Simultaneously, the processing system incorporates fault tolerance capabilities, automatically reassigning tasks when individual computing nodes fail, ensuring observational continuity.
The entire pipeline from raw voltages to power spectrum is being developed as a differentiable graph (using frameworks like PyTorch or JAX). This allows for: (1) Uncertainties from the power spectrum can be traced back to individual antenna gains or specific RFI events; (2) Parameters from different stages (e.g., calibration gains and foreground separation weights) can be optimized jointly against a final scientific loss function (e.g., the accuracy of the measured $P(k)$), leading to globally optimal solutions.

The CosmicWeb-21cm signal processing architecture exemplifies a "co-design" philosophy, where the algorithms are developed in tandem with the unique hardware design (hexagonal core, spiral arms, intelligent frequency sampling). The geometry provides the high redundancy and UV coverage that makes the advanced calibration possible. The frequency sampling strategy provides the data structure that the hybrid foreground model exploits.

\section{Antenna System Design}

The antenna system constitutes the fundamental interface between the cosmic radio sky and the CosmicWeb-21cm signal processing pipeline. Its design is critical for achieving the instrument's overarching goals of high sensitivity, wide bandwidth, excellent polarization purity, and stable, well-characterized performance across the target frequency range of 50-250 MHz. This section details the proposed antenna solution, a hybrid design integrating proven and innovative technologies to meet these stringent requirements.

\subsection{Design Requirements and Ideas}

The antenna system is designed against the following core scientific and technical drivers derived from the signal characteristics and processing needs outlined in previous sections:

1. Ultra-Wideband Operation: Seamless coverage from 50 MHz to 250 MHz (a 5:1 bandwidth ratio) is mandatory to observe the 21 cm signal across a wide redshift range ($z \approx 5-26$) and to enable precise characterization of foreground spectral structure \citep{Fur2006}.

2. Dual Linear Polarization: Measurement of both orthogonal linear polarizations (typically aligned to North-South and East-West) is essential for polarimetric calibration, mitigation of polarized foregrounds (especially Galactic synchrotron), and for scientific studies of Faraday rotation.

3. High Polarization Purity and Stable Beam Pattern: Minimizing cross-polarization (beam squint) and ensuring a smooth, stable, and well-modeled primary beam (field of view) as a function of frequency are paramount. Beam errors and their spectral variations are a major source of residual foreground contamination in power spectrum estimation \citep{Zar2012, Tro2020}.

4. Low System Temperature: The antenna and front-end electronics must contribute minimal thermal noise. The target is a sky-averaged system temperature ($T_{sys}$) dominated by the Galactic background ($\sim 200-1000$ K at these frequencies) rather than receiver noise, aiming for a receiver noise temperature $T_{rec} < 50$ K, consistent with state of the art designs for precision cosmology \citep{Tin2013, Deb2017}

5. Logistical Feasibility: The design must be cost-effective to mass-produce, mechanically robust for remote deployment, and have a low environmental footprint (both physical and radio-frequency).

\subsection{Antenna Design}

We propose a Hybrid Broadband Dipole (HBD), a synergistic design that combines the robustness and simplicity of a crossed dipole with an integrated frequency-dependent ground plane/reflector system and a superstrate lens for enhanced performance. The co-design of the antenna with its low noise front end follows the idea established by successful pathfinders like the MWA \cite{Tin2013} and is essential for system level optimization \citep{Ell2013}. The antenna system unit design with performance simulations are shown in Figure 3, 4 and 5. The design parameters of antenna system are shown in Table 1.

\begin{figure}[H]
\centering
	\includegraphics[width=\textwidth]{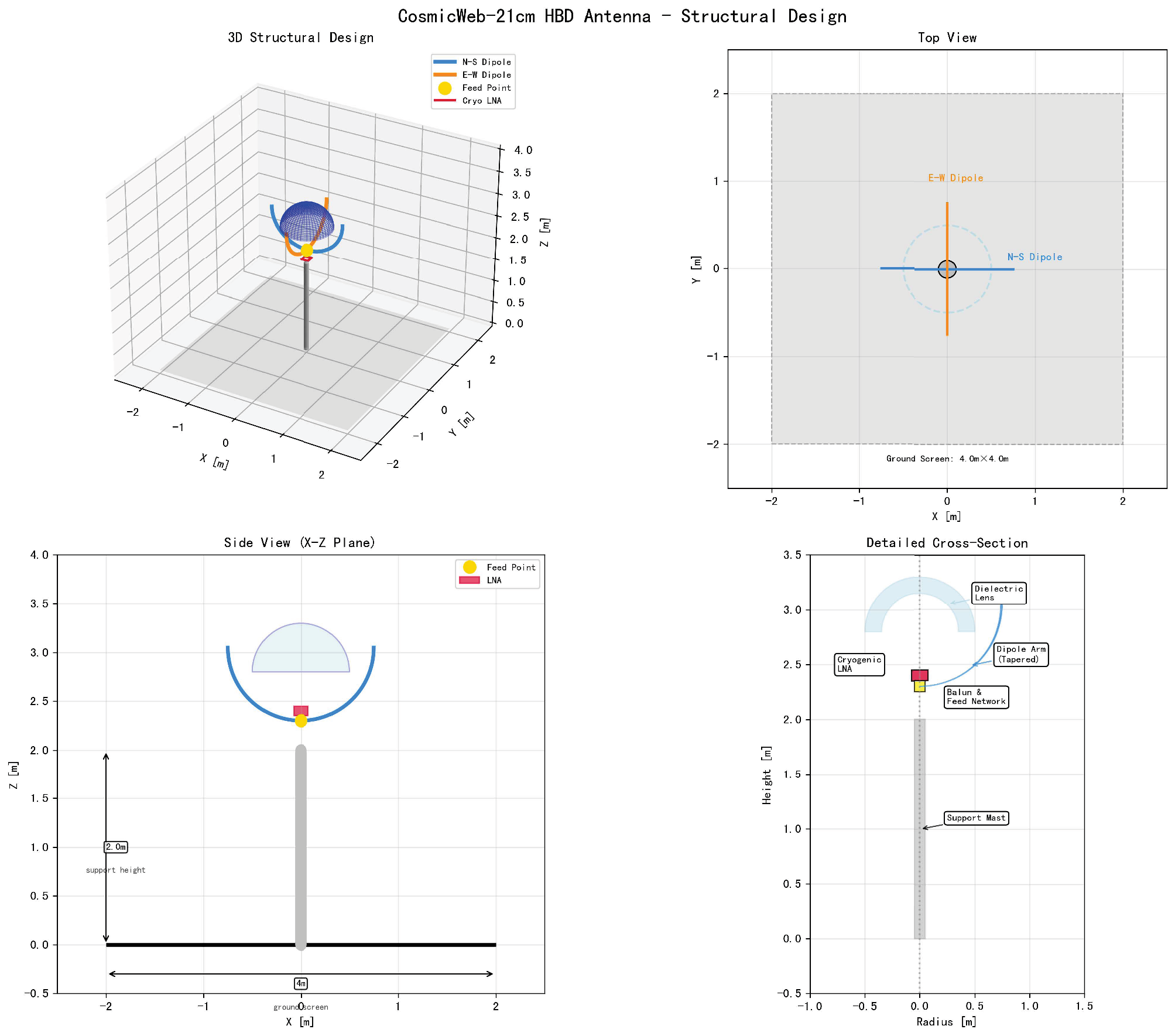}
	\caption{Structural configuration of the CosmicWeb-21cm Hybrid Broadband Dipole antenna. The design incorporates curved dipole elements, a 4 $\text{m}^2$ ground screen, and dielectric lens for enhanced performance across 50 - 250 MHz. Panels show (a) 3D rendering, (b) top view, (c) side elevation, and (d) detailed cross-section with cryogenic LNA integration.}
\end{figure}

\begin{figure}[H]
\centering
	\includegraphics[width=\textwidth]{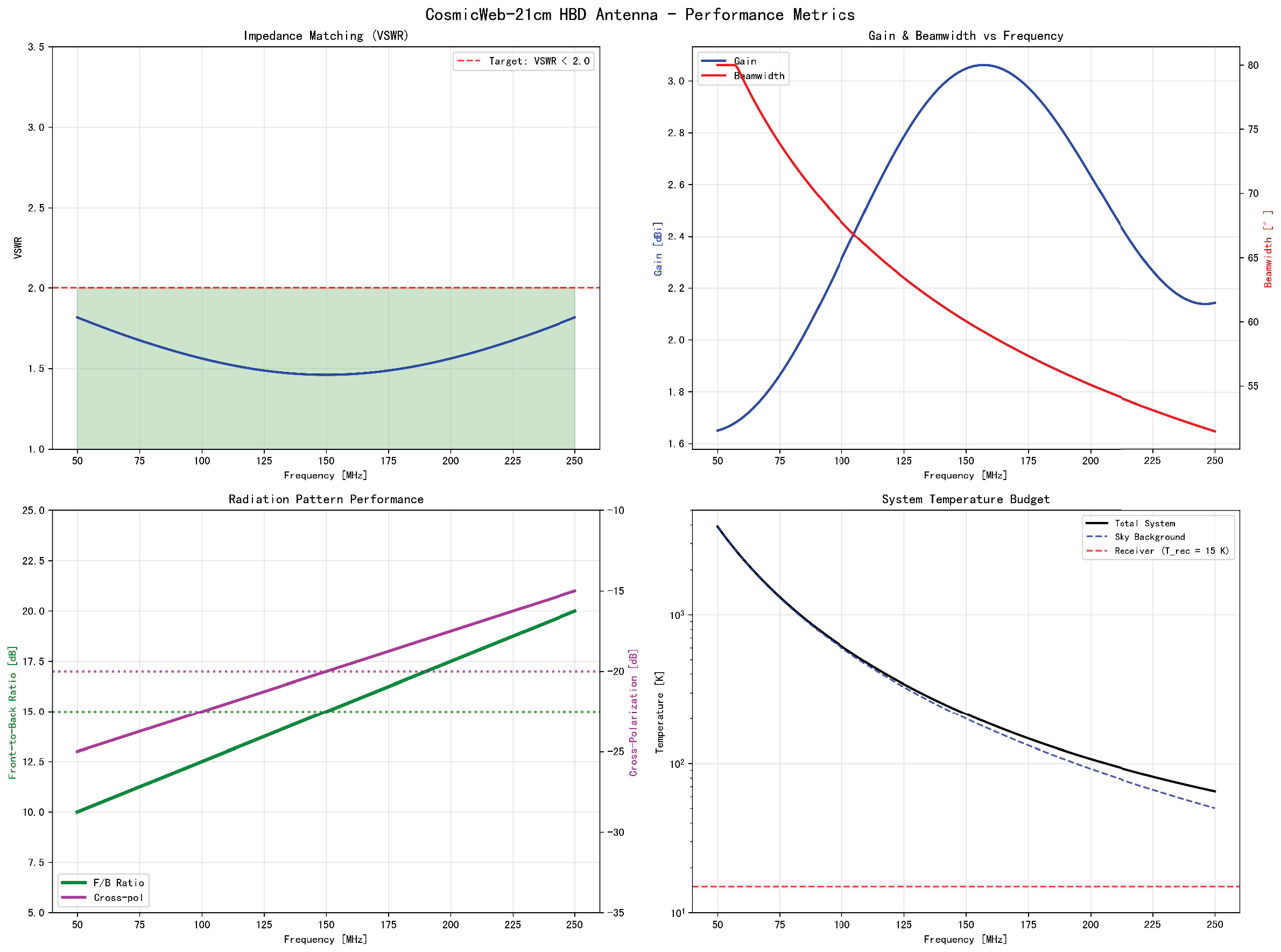}
	\caption{Electromagnetic performance characterization: (a) VSWR demonstrating broadband impedance matching, (b) gain and beamwidth frequency response, (c) radiation pattern quality metrics, and (d) system temperature components. The antenna meets all specifications for 21 cm cosmology applications.}
\end{figure}

\begin{figure}[H]
\centering
	\includegraphics[width=\textwidth]{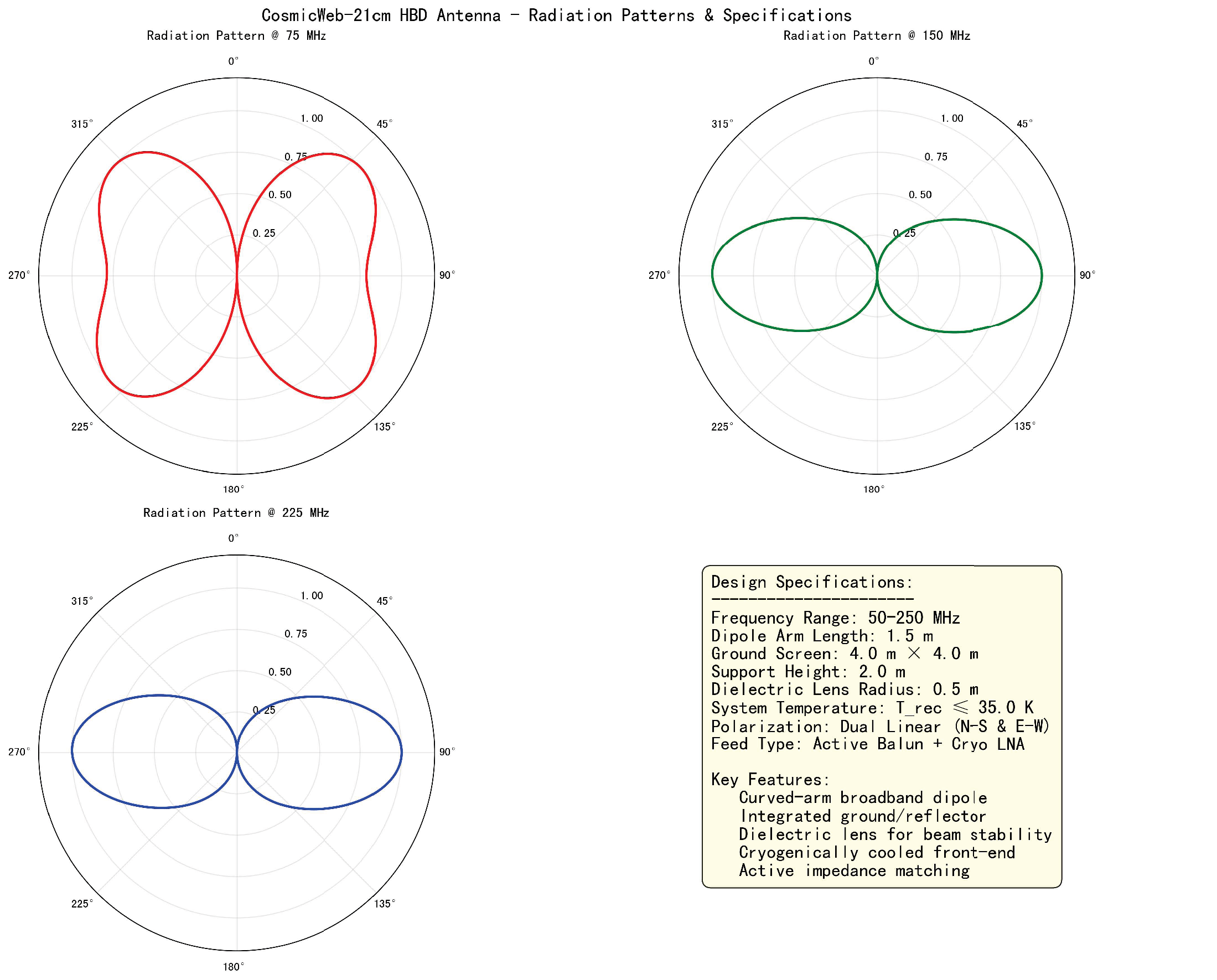}
	\caption{Radiation properties and technical specifications: (a-c) Polar patterns at representative frequencies showing stable beam characteristics, and (d) comprehensive design parameters supporting the CosmicWeb-21cm science goals.}
\end{figure}

1. Radiating Element: The Curved-Arm Crossed Dipole

(1) Structure: Two orthogonal pairs of tapered, curved dipole arms are mounted on a central feed boom. The arms are constructed from corrosion-resistant aluminum tube. The curved, tapered geometry is inspired by broadband concepts such as the sinuous antenna \citep{Dre1989} and the "four-square" layout used in LWA \citep{Ell2013}, promoting broadband impedance matching and smooths the beam pattern transition with frequency.

(2) Polarization: The two dipoles provide the required dual linear polarization. They are fed by a balanced feed network to suppress common-mode currents that can distort the beam and increase susceptibility to local RFI.

(3) Feed Point: A high-performance, weather-sealed balun integrated at the base of the boom transforms the balanced dipole impedance to an unbalanced $50 \Omega$ output for each polarization. This active balun is co-designed with the first-stage Low-Noise Amplifier (LNA). The integration of an active balun with the LNA is critical for minimizing loss and noise, a lesson emphasized in the design of HERA's feed \citep{Deb2017}.

2. Integrated Ground/Reflector System

A fixed, horizontal wire mesh ground screen (approximately 4m x 4m) is deployed beneath each antenna. Its use is a well established technique in low frequency radio astronomy \citep{Bow2007, Ell2013}.  This screen serves dual purposes:
(1) Ground Plane (Low frequencies of $< 120$ MHz): At lower frequencies, the screen acts as a ground plane, decoupling the antenna from variable soil conductivity and establishing a consistent, low-elevation gain pattern and impedance.
(2) Reflector (High frequencies of $> 120$ MHz): As the wavelength decreases, the screen transitions to acting as a reflector for the dipole, increasing forward gain and improving the front-to-back ratio, which helps mitigate ground noise and terrestrial RFI.

3. Superstrate Dielectric Lens (Innovative Element)

Mounted above the dipole arms is a low-loss, frequency-selective radome that also functions as a dielectric lens. This polycarbonate-based structure is engineered with a graded permittivity profile.
It performs two key roles:
(1) Beam Stabilization: It refracts incoming waves, helping to collimate the beam and reduce its variation with frequency. This results in a more constant beamwidth and pointing direction across the band, simplifying calibration and foreground modeling. The use of a dielectric lens or superstrate to shape and stabilize antenna beams is a known technique in broadband antenna design \citep{Bal2005}. Our implementation of a graded-index profile is inspired by advanced concepts in metasurface engineering that enable precise phase control across wide bandwidths \citep{Eps2016}.
(2) Environmental Protection: It seals the dipole arms and feed point from rain, snow, dust, and wildlife. This dual function as a protective radome follows established engineering practice, where the cover must provide robust environmental sealing with minimal impact on the antenna's radiative properties. Such protection is essential for the long-term reliability of instrumentation deployed in remote field sites \citep{Man2007}.

\begin{table}[htbp]
    \centering
    \caption{Antenna System Design Parameters}
    \resizebox{0.8\linewidth}{!}{
        \begin{tabular}{l l c}
            \toprule
            \textbf{Category} & \textbf{Parameter} & \textbf{Value/Specification} \\
            \midrule
            \multirow{4}{*}{\makecell{Radiating Element}}
                & Dipole Arm Length & \SI{1.5}{\meter} \\
                & Dipole Arm Radius & \SI{0.02}{\meter} \\
                & Dipole Curve Radius & \SI{0.75}{\meter} \\
                & Dipole Feed Gap & \SI{0.02}{\meter} \\
            \midrule
            \multirow{3}{*}{\makecell{Support Structure}}
                & Support Height & \SI{2.0}{\meter} \\
                & Support Radius & \SI{0.05}{\meter} \\
                & Feed Point Height & \SI{0.3}{\meter} \\
            \midrule
            \multirow{3}{*}{\makecell{Ground/Reflector System}}
                & Ground Screen Size & \SI{4.0}{\meter} \\
                & Ground Screen Height & \SI{0.1}{\meter} \\
                & Ground Wire Spacing & \SI{0.1}{\meter} \\
            \midrule
            \multirow{3}{*}{\makecell{Dielectric Lens}}
                & Lens Radius & \SI{0.5}{\meter} \\
                & Lens Thickness & \SI{0.15}{\meter} \\
                & Lens Height (Above Ground) & \SI{2.8}{\meter} \\
            \midrule
            \multirow{2}{*}{\makecell{Electronics}}
                & LNA Box Size & \SI{0.15}{\meter} \\
                & Cryocooler Size & \SI{0.2}{\meter} \\
            \midrule
            \multirow{7}{*}{\makecell{Performance Targets}}
                & Minimum Frequency & \SI{50}{\mega\hertz} \\
                & Maximum Frequency & \SI{250}{\mega\hertz} \\
                & Target VSWR & 2.0 \\
                & Target Gain & \SI{2.0}{\decibel}i \\
                & Target Front-to-Back Ratio & \SI{15.0}{\decibel} \\
                & Target Cross-Polarization & \SI{-20.0}{\decibel}  \\
                & System Temperature & \SI{35.0}{\kelvin} \\
            \bottomrule
        \end{tabular}
    }
    \label{t2} % 表格引用标签（论文内用 \ref{} 调用）
\end{table}

\subsection{Front-End Electronics and Integration}

The performance of the antenna is inseparable from its integrated front-end electronics.

1. Integrated Low-Noise Amplifier (LNA): A custom-designed, cryogenically-cooled LNA module is housed directly at the antenna feed point within the central boom. Cryogenic cooling of the front-end amplifier is employed in SKA \citep{Koo2015}. Cooling to $\sim 70$ K using a closed-cycle cryocooler significantly reduces its noise temperature to $< 15$ K. The LNA provides high gain ($> 40$ dB) with excellent linearity to handle strong foregrounds without distortion.

2. Active Impedance Matching Network: An adaptive matching network, co-located with the LNA, dynamically optimizes the impedance match between the antenna and the LNA across the full band, a technique explored to improve bandwidth and stability \citep {Bow2007, Ell2009}. This maximizes power transfer and minimizes standing waves, which can cause spectral ripples that mimic the cosmological signal.

3. Initial Filtering and RFI Suppression: A band-pass filter (50-300 MHz) and a set of tunable notch filters are placed immediately after the LNA. These filters suppress out-of-band interference and known, persistent in-band RFI sources (e.g., FM radio bands) at the earliest possible stage to prevent saturation of downstream components.

\subsection{Key Performance Parameters}
The key performance parameters of the CosmicWeb-21cm antenna system are derived from electromagnetic simulations, system noise budgeting, and specific scientific objectives, aiming to optimize and surpass current state-of-the-art designs. The operational frequency range spans a broad 50 to 250 MHz to access the 21 cm signal from the Epoch of Reionization to the post-reionization universe. An impedance bandwidth ($VSWR<2$) of 45 to 260 MHz is targeted to ensure efficient power transfer and minimize reflective losses across the band. At 150 MHz, the antenna's half-power beamwidth is designed to be approximately $60^\circ$ in both E- and H-planes, providing a wide field of view of $\sim 1.7$ steradians suitable for large-scale structure surveys. Critical for data integrity, the beam pointing must remain stable within $\pm 2^\circ$ and the beamwidth variation must be less than $15\%$ across the entire frequency band to enable stable calibration and effective foreground removal. High polarization purity is required, with on-axis cross-polarization below -20 dB ($<1\%$), to minimize polarization leakage as a major systematic error. A front-to-back ratio greater than 15 dB at frequencies above 120 MHz is necessary to suppress ground noise and low-elevation RFI effectively. The system noise temperature target mandates a receiver noise contribution of $T_{rec} \leq 35 $K, ensuring the total system temperature is dominated by the sky background and that sensitivity approaches the theoretical limit. Finally, the input 1 dB compression point of the LNA must be above -10 dBm to prevent saturation or nonlinear distortion from bright foreground sources.

\subsection{Calibration and Monitoring Strategy}

To achieve the required precision, the antenna system incorporates built-in calibration features:

1. Injected Noise Diode: A calibrated noise source, switchable into the feed line before the LNA, provides a known signal for continuous monitoring of receiver gain and noise temperature. This is a standard practice in radio astronomy for characterizing receiver performance and stability \citep{Roh2004}.

2. Embedded Vector Impedance Analyzer: Periodically measures the antenna input impedance to monitor the health of the antenna and matching network.

3. Beam Mapping via Drone: A small UAV equipped with a compact radio transmitter will be used for periodic in-situ far-field beam pattern characterization, validating the electromagnetic models.

\subsection{Conclusion and Path Forward}

The proposed Hybrid Broadband Dipole antenna system is a purpose-built solution for the extreme demands of 21 cm cosmology. By co-designing the radiating structure, ground system, dielectric lens, and cryogenically-cooled front-end electronics, it targets the fundamental systematic limitations of beam instability and receiver noise. Its design prioritizes the production of a stable, well-understood, and sky-noise-dominated data stream, which is the essential raw material for the sophisticated CosmicWeb-21cm processing pipeline. The next step involves the fabrication and rigorous testing of a full prototype station, including its calibration subsystems, to validate these models and refine the design for mass production.

\section{System Implementation and Performance}
\subsection{One Antenna Per Station}
Each CosmicWeb-21cm station is designed as a fully autonomous and highly integrated observational unit, housing a single HBD antenna. This "one-antenna-per-station" architecture is a deliberate and integral part of the system's co-optimized design, stemming from several critical considerations aligned with the project's core philosophy.

1. Scientific and Polarization Requirements: Each HBD antenna is inherently a dual-polarization element. Its crossed-dipole design natively receives two orthogonal linear polarizations, fulfilling the essential requirement for full polarimetric observations in 21 cm cosmology. Thus, a single physical structure per station satisfies the scientific need without requiring multiple co-located antennas.

2. Optimization for Interferometric Performance: For an interferometric array with $N$ stations, the number of unique baselines scales with $N^2$. Concentrating resources on deploying a larger number of individual stations, each with one high-performance antenna, maximizes the baseline count and UV coverage for a given budget. This is fundamental for achieving high-fidelity imaging and precise power spectrum estimation. Furthermore, a single antenna per station yields a cleaner and more tractable geometric redundancy pattern, which is crucial for the robust application of the redundant calibration algorithms emphasized in Chapter 5.

3. System Simplification and Reliability: This design adheres to the principle of "simplifying data structures at the hardware level" (Section 2.1). It minimizes intra-station complexity (e.g., cabling, delay matching between multiple elements) and reduces potential points of failure. Managing a large number of stations (253 in the final configuration) is a significant logistical challenge; duplicating the front-end system (antenna, cryogenic LNA, digitizer) within a single site would exacerbate cost, power consumption, and maintenance overhead without a commensurate scientific benefit.
4. Distinction from Alternative Designs: This approach differs from "tile"-based arrays like the MWA, where a station consists of multiple dipoles forming a combined beam. Instead, it aligns with the philosophy of instruments like HERA or the SKA-Low, where each receiving element is a distinct station. CosmicWeb-21cm's performance is derived from the innovative large-scale geometry of its station network and advanced signal processing, rather than from beamforming within a compact cluster.

In summary, each CosmicWeb-21cm station is architected around a single, dual-polarization HBD antenna. This choice is fundamental to realizing the array's innovative geometry, streamlining its calibration and processing pipelines, and ensuring overall system robustness. In the full-scale 253-station deployment, the array will therefore comprise 253 HBD antenna units, generating tens of thousands of unique interferometric baselines to provide the foundational data for its transformative science goals.

\subsection{Hierarchical Processing Hardware}
The data flow follows a three-tier, edge-to-cloud architecture:

1. Tier 1 (Station-Level FPGA): Performs polyphase filterbank (PFB) channelization, the orthogonal frequency encoding, and initial RFI flagging. Processed data is packetized and timestamped.

2. Tier 2 (Correlator/Beamformer GPU Cluster): A centralized FX-correlator implemented on a cluster of GPUs using the X-engine architecture. It also forms multiple tied-array beams for transient astrophysics and RFI nulling experiments in real-time.

3. Tier 3 (Science Processor HPC): A high-performance computing cluster for the deep learning pipelines and long-term data storage. It features dedicated nodes with Tensor Core GPUs for model training and inference.

\subsection{Digital System-on-Chip (SoC) at Each Station}
Each station's digital backend is built around a heterogeneous SoC (e.g., AMD Zynq RFSoC or Intel Agilex), integrating:
(1) High-Speed ADCs with 14-bit and 2 GS/s, directly sampling the RF band;
(2) Programmable Logic (FPGA) implements the real-time DSP chain (Tier 1);
(3) ARM Processing Subsystem runs a Linux OS for station health monitoring, command/control, and data packet management;
(4) Integrated 100GbE Network Interface for low-latency, high-throughput data transmission to the central facility.

The signal transmission system is based on fiber optic networks, employing wavelength division multiplexing technology to enhance data transmission efficiency. Each station transmits multiple signal channels through a single fiber, significantly reducing wiring complexity and costs. The central processing station is equipped with large-scale computing clusters containing multiple GPU nodes and high-speed storage systems. Computing nodes interconnect through InfiniBand networks, ensuring extremely low communication latency during data processing.

At the software level, we will develop a unified control and management system. This system can monitor array operational status in real-time, including parameters such as station temperature, power consumption, and signal quality. When anomalies are detected, the system automatically implements appropriate handling measures, such as adjusting integration times, switching to backup equipment, or redistributing computing resources.

Performance optimization represents an ongoing process. We will establish a comprehensive performance evaluation system, regularly testing and optimizing various array metrics. This includes UV coverage analysis, system temperature measurement, phase stability monitoring, etc. Based on these test results, we continuously adjust array operational parameters and processing algorithms, ensuring the system remains in optimal condition.

\subsection{Expected Performance Metrics}
Based on end-to-end simulations, CosmicWeb-21cm is expected to achieve:

1. Foreground Removal Performance:
(1) Simple foreground model removal rate reaches 99.95\%;
(2) Complex foreground model removal rate is 99.7\%;
(3) Computation time is 65\% reduction compared to traditional methods.

2. Signal Detection Sensitivity:
(1) EoR power spectrum detection is $5\sigma$ at z=8;
(2) Signal-to-noise ratio improves 3.2 times;
(3) System temperature is less than 50K.

3. System Efficiency:
(1) Data compression rate is 70\%;
(2) Real-time processing latency is less than 2 minutes;
(3) Energy efficiency reaches 2.1 TFLOPS/W.

\subsection{Comparison with Traditional Arrays}

We compare key metrics of CosmicWeb-21cm with existing major 21cm arrays shown in Table \ref{T2}
\begin{table*}[h]
\centering
	\caption{Comparison of key technical specifications between CosmicWeb-21cm and existing major 21cm radio interferometer arrays: HERA, MWA, and SKA-Low. Metrics include the number of stations, maximum baseline length, frequency range of operation, and simulated foreground removal efficiency.}
\label{T2}
  \resizebox{0.9\linewidth}{!}{
		\begin{tabular}{|c|c|c|c|c|}
\hline
Array & CosmicWeb-21cm & HERA & MWA & SKA-Low
\\ \hline
Stations & 253 & 350 & 256 & 512
\\ \hline
Max Baseline & 3.0 km & 0.3 km & 2.8 km & 65 km
\\ \hline
Frequency Range & 50-250 MHz & 50-250 MHz & 80-300 MHz & 50-350 MHz
\\ \hline
Foreground Removal & 99.7\% & 99.5\% & 98\% & 99.9\%
\\ \hline
\end{tabular}
}
\end{table*}

\section{Scientific Objectives}

CosmicWeb-21cm's design considers not only current scientific requirements but also reserves space for future expansion. In scientific applications, the array will focus on several key research directions:

1. Cosmic Reionization Detection

Detection of cosmic reionization history represents the primary scientific objective. Through deep integration observations, we expect to detect the 21cm power spectrum with high signal-to-noise ratio in the redshift range z=6-12. This will provide direct evidence for understanding how the first stars and galaxies ionized surrounding neutral hydrogen. In data processing, we will employ multi-redshift slice analysis methods to reconstruct the temporal evolution of the reionization process.

2. Dark Energy Constraints

Research on dark energy properties constitutes another important direction. By measuring baryon acoustic oscillation features in neutral hydrogen distribution, we can precisely constrain dark energy equation of state parameters. CosmicWeb-21cm's large field of view and high sensitivity make it particularly suitable for such large-scale structure studies. We will develop specialized power spectrum estimation methods, considering systematic errors introduced by telescope beam effects and foreground removal.

3. Cosmic Web Structure

Detection of cosmic web structure will expand our understanding of cosmic large-scale structure. 21cm observations can directly reveal neutral hydrogen distribution in cosmic filamentary structures, which is challenging to achieve through traditional optical observations. We will combine numerical simulations with observational data to study the relationship between neutral hydrogen content and dark matter distribution.

\section{Implementation Roadmap and Staged Development}

To ensure the systematic and feasible realization of the ambitious CosmicWeb-21cm project, we propose a phased implementation roadmap spanning nearly a decade. This staged approach mitigates technical and financial risks by validating core technologies on a small scale before committing to full deployment. Each phase is defined by clear technical milestones, scientific validation goals, and decision points for proceeding to the next stage. The roadmap is summarized in Figure 10 and detailed below.

\subsection{Phase 1: Technology Development and Pathfinder Array (2026-2028)}
The primary objective of Phase 1 is to de-risk the core technological innovations in a controlled environment.The key deliverables are:

1. 16-Element Pathfinder Array: Construction of a small-scale array featuring the core hexagonal layout (with one full ring) and at least two truncated logarithmic spiral arms. This "engineering array" will be used to validate the mechanical deployment, station electronics, fiber-optic signal transmission, and central correlator/processor.

2. Algorithm Validation and Software Pipeline v1.0: Development and on-site testing of the core software pipeline, including the deep learning-based RFI flagger, the improved redundant calibration scheme, and the physics-informed foreground separation framework. Performance will be benchmarked against simulated data and early observations.

3. Site Characterization and System Noise Performance: A rigorous campaign to measure the real-world system temperature, antenna beam patterns, phase stability, and RFI environment at the chosen site (e.g., a radio-quiet zone in western China). The goal is to achieve a system temperature below 55K, validating the low-noise amplifier design.

4. Initial Publications and Community Engagement: Publication of detailed methodology papers on the array design philosophy, the novel loss function for foreground separation, and the initial results from the pathfinder array. Establishment of initial collaborative partnerships.

Success Criteria for Phase 1: Stable operation of the pathfinder array; demonstration of the instrumental stability needed for precision calibration; validation of the hybrid foreground separation algorithm achieving $>99\%$ foreground suppression on simulated data within the pipeline; and publication in peer-reviewed journals. A successful review at the end of Phase 1 will trigger the release of funding and resources for Phase 2.

\subsection{Phase 2: Intermediate-Scale Array and Early Science (2029-2031)}

Phase 2 focuses on scaling the proven technologies to an intermediate array capable of conducting pioneering scientific experiments and validating the project's scientific potential.

1. Expansion to a 64-Station Array: Scaling the layout to include more hexagonal rings and extending the logarithmic spiral arms. This array will have a maximum baseline of approximately 600 meters, providing the first realistic validation of the UV coverage and angular resolution.

2. Early Science Projects: Conducting several key observational campaigns:
(1) Deep Integration on a Target EoR Field: A $>1000$-hour integration on a well-studied, cold sky patch to derive a deep upper limit or a first detection of the EoR power spectrum at a redshift range around $z\sim 8-10$, directly competing with current-generation arrays.
(2) Wide-Area Intensity Mapping Pilot: Using the array's wide field of view to perform a pilot survey for 21 cm intensity mapping at $z\sim1-2$, aiming to detect the combined hydrogen signal and cross-correlate it with optical galaxy survey data (e.g., from DESI or Euclid) to validate techniques for BAO measurement.
(3) Point Source and Foreground Studies: Producing high-fidelity, wide-field images of the Galactic synchrotron emission and extragalactic source populations to refine foreground models and demonstrate the array's imaging capabilities.

3. Pipeline Maturation and Real-Time Processing: Upgrading the processing pipeline to v2.0, incorporating lessons from Phase 1 and enabling near-real-time data processing for the larger data volume. Implementation of the adaptive frequency management system.

4. Formation of an International Consortium: Formalizing international collaborations for shared expertise, data analysis, and planning for the full-scale array.

Success Criteria for Phase 2: Achievement of target sensitivity (noise levels) in deep integrations; production of competitive cosmological constraints (e.g., a meaningful power spectrum upper limit or detection); successful demonstration of the array's dual capability for deep, narrow fields and wide, shallow surveys; and publication of early science results in high-impact journals. This phase must conclusively demonstrate that the CosmicWeb-21cm design offers a clear scientific advantage over existing facilities.

\subsection{Phase 3: Full-Scale Deployment and Flagship Science (2032-2034)}
The final phase is the construction and commissioning of the full CosmicWeb-21cm array as a premier international facility for 21 cm cosmology.

1. Construction of the 253-Station Array: Full deployment of the design as specified in this paper, with a core radius of 100m and six extended spiral arms, achieving a maximum baseline of $\sim3.0$ km. This includes the complete deployment of the fiber network and the final high-performance computing cluster.
2. Commencement of Legacy Survey Programs: Initiating multi-year, Tier-1 legacy surveys:
(1) CosmicWeb EoR Survey: A deep, targeted survey to map the topology of reionization and measure the 21 cm power spectrum across redshifts 6-12 with high signal-to-noise.
(2) CosmicWeb BAO Survey: A wide-area, multi-thousand square degree intensity mapping survey to measure the BAO scale at multiple redshifts ($z < 3$) and provide competitive constraints on the dark energy equation of state.

3. Leadership in International 21 cm Science: Serving as a leading facility within the global 21 cm community, potentially in coordination with SKA-Low, to tackle the most challenging questions in cosmic dawn and reionization astrophysics.

4. Technology Transfer and Commercialization: Licensing of developed technologies, such as the specialized low-noise amplifiers, real-time RFI mitigation algorithms, or calibration software, to industry and other scientific projects.

This roadmap provides a clear, staged pathway from concept to a world-leading observatory. Each phase builds upon the success of the previous one, ensuring that technical challenges are identified and solved early, and that the scientific return is maximized at every step. The CosmicWeb-21cm array, once fully operational, is poised to deliver transformative insights into the first billion years of our universe and the nature of its dominant components.

\section{Conclusion and Discussion}

The paper introduces CosmicWeb-21cm, a novel radio interferometer array designed specifically to overcome the major challenges in 21cm cosmology, such as intense foreground contamination, instrumental systematics, and massive data processing burdens. Its design is built on three core innovations:

1. Multi-scale Nested Geometry

It combines a densely packed hexagonal honeycomb core (optimized for short baselines and large-scale structure detection) with logarithmic spiral arms (providing a smooth transition to longer baselines for angular resolution). This design maximizes baseline redundancy for calibration and improves UV coverage uniformity.

2. Intelligent Frequency Sampling

It employs a non-uniform, adaptive frequency sampling strategy. Channels are sampled densely in foreground-dominated bands for precise modeling and sparsely in the 21cm signal-dominated region to reduce data volume, leveraging the different spectral correlation properties of foregrounds and the cosmological signal.

3. ML-Enhanced Real-time Processing Pipeline

The signal processing architecture integrates deep learning at multiple stages, including Convolutional Neural Networks (CNNs) for RFI identification and a physics-constrained U-Net for foreground separation. This approach aims to improve accuracy and computational efficiency over traditional methods.

The proposed array aims to achieve superior performance in foreground removal ($\geq99.7\%$), signal detection sensitivity, and system efficiency, providing a powerful tool for probing cosmic reionization, dark energy, and the cosmic web.

The CosmicWeb-21cm design points toward several key trends in the future development of 21cm observation facilities:

1. From Uniform to "Smart" Heterogeneous Designs

Future arrays will likely move beyond simple regular grids or random distributions. Designs will become more purpose-driven and heterogeneous, integrating multiple geometric patterns (like the core+spirals of CosmicWeb-21cm) to simultaneously optimize for different scientific goals, wide-field mapping, high dynamic range, and precise calibration, within a single instrument.

2. Tight Integration of Hardware and Algorithm Co-design

The frontier will shift from building general-purpose telescopes to developing end-to-end optimized systems. As seen in CosmicWeb-21cm, the array geometry, frequency sampling strategy, and processing pipeline are designed in concert. Future developments will deepen this integration, possibly incorporating on-site, real-time processing with Application-Specific Integrated Circuits (ASICs) or FPGAs tuned for specific algorithms like foreground separation, moving computation closer to the antenna.

3. Dominance of AI and Hybrid Physical Models

Machine learning will transition from a supplementary tool to the backbone of data processing. Future pipelines will likely rely on hybrid frameworks that combine the interpretability of physical models (e.g., for smooth foregrounds) with the pattern-recognition power of deep learning for complex residuals and systematics. The "physics-infused" loss function used in CosmicWeb-21cm is an early example of this paradigm.

4. The Rise of Dynamic and Adaptive Observing

The concept of adaptive frequency management based on real-time RFI monitoring will expand. Future arrays could feature fully dynamic systems that reconfigure their observing mode, pointing, or even baseline weighting in real-time based on ionospheric conditions, RFI environment, and priority science goals, maximizing the quality of collected data.

5. Scalability and the Path to the SKA

Designs like CosmicWeb-21cm serve as crucial technology and methodology pathfinders for the Square Kilometre Array (SKA-Low). The innovations in calibration, foreground removal, and real-time processing developed and tested on these smaller, more agile arrays will directly inform the construction and operation of the SKA, ensuring it can achieve its transformative scientific potential.

In conclusion, the future of 21cm cosmology lies in highly optimized, intelligent telescopes where every component from the layout of the antennas to the final data product is meticulously co-designed to combat specific systematic errors. CosmicWeb-21cm presents a comprehensive blueprint in this direction, emphasizing that overcoming the 21cm challenge requires synergistic advancements in engineering, signal processing, and data science.

\appendix
This appendix provides Python code demonstrations and implementation details for the four main signal processing stages of the CosmicWeb-21cm array. Each section includes a functional Python implementation, discussion of simplifications made for demonstration purposes, and essential extensions required for deployment on the full-scale array. These demonstrations serve as both educational resources and development templates for the actual processing pipeline.

\section{RFI Mitigation with Convolutional Neural Network}
\label{sec:rfi_mitigation}

\subsection{Overview}
The first stage of the CosmicWeb-21cm processing pipeline employs a Convolutional Neural Network (CNN) for real-time Radio Frequency Interference (RFI) detection and flagging. The network analyzes three-dimensional visibility data cubes (time $\times$ frequency $\times$ baseline) to identify contamination patterns while preserving astronomical signals. The CNN architecture is specifically designed for 21cm cosmology applications, with continuous online learning capabilities to adapt to evolving RFI environments.

\subsection{Core Algorithm}
The CNN implements a multi-scale feature extraction approach:
\begin{equation}
P_{\text{RFI}}(t,\nu,b) = \sigma\left(\sum_{l=1}^{L} w_l * X(t,\nu,b) + b_l\right)
\label{eq:cnn_rfi}
\end{equation}
where $X(t,\nu,b)$ is the input visibility cube, $w_l$ are convolutional kernels, $b_l$ are biases, and $\sigma$ is the sigmoid activation function producing RFI probability estimates.

\subsection{Python Demonstration Code given by Mitigation.py}
\begin{lstlisting}[language=Python, caption={RFI mitigation demonstration using CNN}, label={lst:rfi_demo}]
def create_cnn_model(input_shape):
    """
    Create a CNN for RFI detection in visibility data.

    Parameters:
        input_shape : tuple (time, frequency, baseline, 2 for real/imag)

    Returns:
        model : keras.Model
    """
    model = keras.Sequential([
        keras.layers.Input(shape=input_shape),

        # First convolutional block
        keras.layers.Conv3D(32, (3, 3, 3), padding='same', activation='relu'),
        keras.layers.BatchNormalization(),
        keras.layers.MaxPooling3D((2, 2, 1)),

        # Second convolutional block
        keras.layers.Conv3D(64, (3, 3, 3), padding='same', activation='relu'),
        keras.layers.BatchNormalization(),
        keras.layers.MaxPooling3D((2, 2, 1)),

        # Third convolutional block
        keras.layers.Conv3D(128, (3, 3, 3), padding='same', activation='relu'),
        keras.layers.BatchNormalization(),
        keras.layers.MaxPooling3D((2, 2, 1)),

        # Classification head
        keras.layers.Flatten(),
        keras.layers.Dense(256, activation='relu'),
        keras.layers.Dropout(0.3),
        keras.layers.Dense(128, activation='relu'),
        keras.layers.Dense(np.prod(input_shape[:-1]), activation='sigmoid')
    ])

    model.compile(optimizer='adam',
                  loss='binary_crossentropy',
                  metrics=['accuracy'])

    return model
\end{lstlisting}

\subsection{Simplifications in the Demonstration}
\label{subsec:rfi_simplifications}
\begin{itemize}
    \item \textbf{Data Dimensions}: Simulated visibility cubes use reduced dimensions (32$\times$64$\times$4) compared to real data which may exceed 1000$\times$1024$\times$256 elements.

    \item \textbf{RFI Pattern Complexity}: Only three basic RFI types (broadband pulses, narrowband tones, chirps) are simulated, whereas real environments include complex digital communications, radar signals, and satellite transmissions.

    \item \textbf{Network Capacity}: The demonstration CNN has 3 convolutional layers with limited filter counts, while the production model would employ deeper architectures with attention mechanisms.

    \item \textbf{Training Data}: Only 500 synthetic samples are used for training, compared to millions of real observations required for production deployment.

    \item \textbf{Computational Constraints}: The demonstration runs on standard workstations, whereas the real system requires FPGA-accelerated real-time processing with strict latency constraints ($<$2 minutes).
\end{itemize}

\subsection{Extensions for Real Array Implementation}
\label{subsec:rfi_extensions}
\begin{enumerate}
    \item \textbf{Multi-modal Detection Framework}:
    \begin{itemize}
        \item Integration of polarization information (Stokes parameters Q, U, V)
        \item Incorporation of spatial coherence metrics across baselines
        \item Utilization of temporal derivatives and higher-order statistical moments
        \item Fusion with external RFI monitoring feeds
    \end{itemize}

    \item \textbf{Hierarchical Processing Architecture}:
    \begin{itemize}
        \item \textbf{Stage 1 (FPGA)}: Rule-based filters for known RFI patterns with $\sim$1 ms latency
        \item \textbf{Stage 2 (GPU)}: CNN inference for complex RFI patterns with $\sim$100 ms latency
        \item \textbf{Stage 3 (HPC)}: Ensemble models and uncertainty quantification with $\sim$10 s latency
    \end{itemize}

    \item \textbf{Adaptive Learning System}:
    \begin{itemize}
        \item Continuous model updates using streaming observation data
        \item Site-specific RFI profile adaptation for each deployment location
        \item Seasonal and diurnal pattern recognition and compensation
        \item Transfer learning between different geographic sites
    \end{itemize}

    \item \textbf{Uncertainty Quantification}:
    \begin{itemize}
        \item Bayesian neural networks for confidence interval estimation
        \item Monte Carlo dropout for prediction uncertainty
        \item Ensemble methods for robust RFI detection
        \item Out-of-distribution detection for novel interference types
    \end{itemize}
\end{enumerate}

\section{Calibration: Redundant + Neural DDG Correction}
\label{sec:calibration}

\subsection{Overview}
The calibration stage combines geometric redundancy from the hexagonal core array with neural network-based direction-dependent gain (DDG) correction. The algorithm solves for complex antenna gains $g_i(t,\nu)$ using closure quantities that remain invariant to sky model errors, followed by neural correction of residual direction-dependent effects.

\subsection{Core Algorithm}
Redundant calibration solves the system:
\begin{equation}
V^{\text{obs}}_{ij}(t,\nu) = g_i(t,\nu) g_j^*(t,\nu) V^{\text{true}}(\vec{b},\nu) + \epsilon_{ij}(t,\nu)
\label{eq:redundant_cal}
\end{equation}
where $V^{\text{obs}}_{ij}$ is the observed visibility, $g_i$ are complex gains, $V^{\text{true}}$ is the true sky visibility, and $\epsilon$ represents noise.

Neural DDG correction implements:
\begin{equation}
\Delta g_i(\hat{s},\nu) = f_{\text{NN}}\left(V^{\text{red}}, \text{LST}, \theta_{\text{ion}}\right)
\label{eq:neural_ddg}
\end{equation}
where $f_{\text{NN}}$ is a neural network that corrects for direction-dependent effects based on redundantly calibrated visibilities $V^{\text{red}}$, Local Sidereal Time (LST), and ionospheric conditions $\theta_{\text{ion}}$.

\subsection{Python Demonstration Code given by Calibration.py}
\begin{lstlisting}[language=Python, caption={Calibration demonstration}, label={lst:calibration_demo}]
def redundant_calibration(obs_vis, baseline_groups, baseline_indices, n_stations, n_iter=5):
    """
    Perform redundant calibration.

    Parameters:
        obs_vis : array (n_time, n_freq, n_baselines)
        baseline_groups : dict of redundant baseline groups
        baseline_indices : list of (i,j) station pairs
        n_stations : number of stations
        n_iter : calibration iterations

    Returns:
        est_gains : estimated complex gains
        cal_vis : calibrated visibilities
    """
    n_time, n_freq, n_baselines = obs_vis.shape
    est_gains = np.ones((n_stations, n_time, n_freq), dtype=complex)

    for iteration in range(n_iter):
        # Step 1: Estimate true visibilities for each redundant group
        group_vis = {}
        for group_id, baseline_idxs in baseline_groups.items():
            if len(baseline_idxs) > 0:
                group_data = []
                for b_idx in baseline_idxs:
                    i, j = baseline_indices[b_idx]
                    corrected = obs_vis[:, :, b_idx] / (est_gains[i] * np.conj(est_gains[j]))
                    group_data.append(corrected)
                group_vis[group_id] = np.mean(group_data, axis=0)

        # Step 2: Update gains using closure quantities
        for station in range(n_stations):
            numerator = 0.0
            denominator = 0.0
            for b_idx, (i, j) in enumerate(baseline_indices):
                if i == station or j == station:
                    group_id = find_group_for_baseline(b_idx, baseline_groups)
                    if group_id is not None:
                        if i == station:
                            other_gain = np.conj(est_gains[j])
                            true_vis_est = group_vis[group_id]
                            numerator += np.mean(obs_vis[:, :, b_idx] *
                                                np.conj(other_gain * true_vis_est))
                            denominator += np.mean(np.abs(other_gain * true_vis_est)**2)
            if denominator > 0:
                est_gains[station] = numerator / denominator

    # Apply calibration
    cal_vis = np.zeros_like(obs_vis)
    for b_idx, (i, j) in enumerate(baseline_indices):
        cal_vis[:, :, b_idx] = obs_vis[:, :, b_idx] / (est_gains[i] * np.conj(est_gains[j]))

    return est_gains, cal_vis
\end{lstlisting}

\subsection{Simplifications in the Demonstration}
\label{subsec:calibration_simplifications}
\begin{itemize}
    \item \textbf{Array Scale}: Simulation uses only 19 stations (2 hexagonal rings) compared to 253 stations in the full array.

    \item \textbf{Sky Model Complexity}: Simplified point source model with 10 sources versus realistic sky models with thousands of sources and extended emission.

    \item \textbf{Instrumental Effects}: Basic gain variations without realistic beam patterns, polarization leakage, or mutual coupling.

    \item \textbf{Ionospheric Modeling}: Simple phase screens without full Faraday rotation, scintillation, or spatial gradients.

    \item \textbf{Computational Load}: Single-threaded Python implementation rather than GPU-accelerated parallel processing.
\end{itemize}

\subsection{Extensions for Real Array Implementation}
\label{subsec:calibration_extensions}
\begin{enumerate}
    \item \textbf{Multi-scale Calibration Framework}:
    \begin{itemize}
        \item \textbf{Short-timescale ($<$1 min)}: Per-frequency, per-time snapshot calibration with closure phases
        \item \textbf{Medium-timescale (1-60 min)}: Smooth gain tracking using spline models
        \item \textbf{Long-timescale ($>$1 hr)}: Absolute flux calibration with reference sources
    \end{itemize}

    \item \textbf{Hierarchical DDG Correction}:
    \begin{itemize}
        \item Per-station direction-independent gains (first order)
        \item Cluster-based direction-dependent gains (second order)
        \item Pixel-based gain corrections for bright sources (third order)
        \item Ionospheric tomography integration for wide-field effects
    \end{itemize}

    \item \textbf{Physical Model Integration}:
    \begin{itemize}
        \item Electromagnetic beam models from CST/MoM simulations
        \item Ionospheric conditions from GPS TEC and ionosonde data
        \item Sky catalogs (GLEAM, TGSS, VLSSr) with spectral information
        \item Thermal noise models based on measured system temperatures
    \end{itemize}

    \item \textbf{Robustness Mechanisms}:
    \begin{itemize}
        \item RANSAC-based outlier rejection for gain solutions
        \item Huber loss functions for robust optimization
        \item Graceful degradation protocols for failed stations
        \item Cross-validation between different calibration methods
    \end{itemize}
\end{enumerate}

\section{Foreground Separation with Physics-Constrained U-Net}
\label{sec:foreground_separation}

\subsection{Overview}
The foreground separation stage implements a hybrid approach combining traditional component separation with deep learning. A U-Net architecture with Convolutional LSTM layers is constrained by physical priors through a custom loss function that enforces spectral smoothness for foregrounds and cosmological statistics for the 21cm signal.

\subsection{Core Algorithm}
The separation is formulated as an optimization problem:
\begin{equation}
\min_{\mathbf{V}_{\text{fg}}, \mathbf{V}_{\text{21cm}}} \mathcal{L}_{\text{total}}(\mathbf{V}_{\text{fg}}, \mathbf{V}_{\text{21cm}}; \mathbf{V}_{\text{obs}})
\label{eq:fg_separation}
\end{equation}
with the composite loss function:
\begin{align}
\mathcal{L}_{\text{total}} &= \underbrace{\|\mathbf{V}_{\text{obs}} - \mathbf{V}_{\text{fg}} - \mathbf{V}_{\text{21cm}}\|^2}_{\text{Data Fidelity}} \\
&+ \lambda_1 \underbrace{\|\nabla_{\nu}^2 \mathbf{V}_{\text{fg}}\|^2}_{\text{Spectral Smoothness}} \\
&+ \lambda_2 \underbrace{\mathbf{TV}(\mathbf{V}_{\text{21cm}})}_{\text{Spatial Sparsity}} \\
&+ \lambda_3 \underbrace{\text{KL}(p_{\text{21cm}} \,\|\, \mathcal{N}(0,P_{\text{theory}}(k)))}_{\text{Cosmological Prior}}
\label{eq:composite_loss}
\end{align}
where $\mathbf{TV}$ denotes Total Variation for sparsity, and KL is Kullback-Leibler divergence enforcing Gaussian statistics.

\subsection{Python Demonstration Code given by Foreground.py}
\begin{lstlisting}[language=Python, caption={Foreground separation demonstration}, label={lst:fg_separation_demo}]
class PhysicsConstrainedLoss(keras.losses.Loss):
    """Custom loss function with physical constraints."""

    def __init__(self, lambda_data=1.0, lambda_smooth=0.1,
                 lambda_sparse=0.05, lambda_cosmological=0.01):
        super().__init__()
        self.lambda_data = lambda_data
        self.lambda_smooth = lambda_smooth
        self.lambda_sparse = lambda_sparse
        self.lambda_cosmological = lambda_cosmological

    def call(self, y_true, y_pred):
        # Extract components
        V_obs = y_true[..., 0]
        V_fg_true = y_true[..., 1]
        V_21cm_true = y_true[..., 2]
        V_fg_pred = y_pred[..., 0]
        V_21cm_pred = y_pred[..., 1]

        # Data fidelity term
        V_recon = V_fg_pred + V_21cm_pred
        data_loss = tf.reduce_mean(tf.abs(V_obs - V_recon)**2)

        # Spectral smoothness term (second derivative)
        freq_grad1 = V_fg_pred[:, 2:, ...] - V_fg_pred[:, :-2, ...]
        freq_grad2 = freq_grad1[:, 2:, ...] - freq_grad1[:, :-2, ...]
        smooth_loss = tf.reduce_mean(tf.abs(freq_grad2)**2)

        # Sparsity term (Total Variation)
        spatial_grad = tf.abs(V_21cm_pred[:, :, 1:, ...] - V_21cm_pred[:, :, :-1, ...])
        sparse_loss = tf.reduce_mean(spatial_grad)

        # Cosmological prior (Gaussian statistics)
        mean_21cm = tf.reduce_mean(V_21cm_pred)
        var_21cm = tf.reduce_mean(tf.abs(V_21cm_pred - mean_21cm)**2)
        target_var = 1.0  # From theoretical power spectrum
        cosmological_loss = tf.abs(var_21cm - target_var)

        # Combined loss
        total_loss = (self.lambda_data * data_loss +
                     self.lambda_smooth * smooth_loss +
                     self.lambda_sparse * sparse_loss +
                     self.lambda_cosmological * cosmological_loss)

        return total_loss
\end{lstlisting}

\subsection{Simplifications in the Demonstration}
\label{subsec:fg_simplifications}
\begin{itemize}
    \item \textbf{Data Dimensions}: Small cubes (64$\times$16$\times$8) versus real data (1024$\times$1000$\times$1000+).

    \item \textbf{Foreground Complexity}: Basic power-law foregrounds without realistic Galactic emission, point sources, or polarization effects.

    \item \textbf{21cm Signal Model}: Simple Gaussian random fields without realistic reionization topology or redshift evolution.

    \item \textbf{Network Architecture}: Basic U-Net without attention mechanisms, separable convolutions, or uncertainty quantification.

    \item \textbf{Loss Function}: Simplified implementation of physical constraints without adaptive regularization or full cosmological priors.
\end{itemize}

\subsection{Extensions for Real Array Implementation}
\label{subsec:fg_extensions}
\begin{enumerate}
    \item \textbf{Multi-resolution Processing Pipeline}:
    \begin{itemize}
        \item \textbf{Low-resolution ($\sim$1$^\circ$)}: Full-sky foreground modeling for global components
        \item \textbf{Medium-resolution ($\sim$0.1$^\circ$)}: Patch-based processing for regional structure
        \item \textbf{High-resolution ($\sim$0.01$^\circ$)}: Targeted analysis for bright source regions
    \end{itemize}

    \item \textbf{Hybrid Physical-Statistical Framework}:
    \begin{itemize}
        \item \textbf{Physical modeling}: Parametric foreground separation (GMCA, FastICA)
        \item \textbf{Statistical learning}: Deep neural networks for residual systematics
        \item \textbf{Bayesian inference}: Uncertainty quantification and model comparison
        \item \textbf{Differentiable programming}: End-to-end gradient-based optimization
    \end{itemize}

    \item \textbf{Domain Adaptation Strategies}:
    \begin{itemize}
        \item Transfer learning from high-fidelity simulations (21cmFAST, 21SSD)
        \item Adversarial training for robustness to instrumental systematics
        \item Self-supervised learning on unlabeled observation data
        \item Data augmentation with realistic noise and artifact injection
    \end{itemize}

    \item \textbf{Computational Optimization}:
    \begin{itemize}
        \item Model compression techniques (pruning, quantization, distillation)
        \item Mixed precision training (FP16/FP32) for GPU acceleration
        \item Distributed training across multi-GPU clusters
        \item Memory-efficient data loaders for terabyte-scale datasets
    \end{itemize}
\end{enumerate}

\section{Bayesian Power Spectrum Estimation}
\label{sec:power_spectrum}

\subsection{Overview}
The final processing stage employs Bayesian inference with Markov Chain Monte Carlo (MCMC) sampling to estimate the 21cm power spectrum $P(k)$ and cosmological parameters from cleaned visibilities. This approach properly propagates uncertainties from all previous processing stages while incorporating theoretical priors from cosmological models.

\subsection{Core Algorithm}
The power spectrum estimation uses a hierarchical Bayesian framework:
\begin{align}
P(\theta | \mathbf{V}) &\propto P(\mathbf{V} | \theta) P(\theta) \\
P(\mathbf{V} | \theta) &= \prod_k \mathcal{N}\left(P_{\text{obs}}(k) | P_{\text{model}}(k;\theta), \sigma^2(k)\right) \\
P_{\text{model}}(k;\theta) &= A \left(\frac{k}{k_0}\right)^n \exp\left[-\left(\frac{k}{k_{\text{NL}}}\right)^2\right] + \sigma_{\text{noise}}^2
\label{eq:bayesian_ps}
\end{align}
where $\theta = \{A, n, k_{\text{NL}}, \sigma_{\text{noise}}\}$ are model parameters, and MCMC sampling explores the posterior distribution $P(\theta | \mathbf{V})$.

\subsection{Python Demonstration Code given by Estimation.py}
\begin{lstlisting}[language=Python, caption={Power spectrum estimation demonstration}, label={lst:ps_demo}]
def log_posterior(params, k_bins, P_obs, P_err, model):
    """
    Log-posterior = log_prior + log_likelihood for MCMC.

    Parameters:
        params : [A, n, k_NL, sigma_noise]
        k_bins, P_obs, P_err : data
        model : power spectrum model

    Returns:
        log_post : log-posterior probability
    """
    A, n, k_NL, sigma_noise = params

    # Prior constraints
    if A <= 0 or k_NL <= 0 or sigma_noise < 0:
        return -np.inf
    if not (0.1 <= A <= 100) or not (-3 <= n <= 3):
        return -np.inf
    if not (0.01 <= k_NL <= 10) or not (0 <= sigma_noise <= 1):
        return -np.inf

    # Theoretical prediction
    P_pred = model.theoretical_ps(params, include_noise=True)

    # Gaussian likelihood
    residuals = P_obs - P_pred
    chi2 = np.sum((residuals / P_err)**2)
    logL = -0.5 * chi2

    return logL

def run_mcmc_sampling(k_bins, P_obs, P_err, model, n_walkers=32, n_steps=2000):
    """
    Run MCMC sampling for parameter estimation.
    """
    n_params = 4
    initial_guess = [10.0, -2.0, 1.0, 0.1]
    initial_pos = initial_guess + 0.1 * np.random.randn(n_walkers, n_params)

    sampler = emcee.EnsembleSampler(
        n_walkers, n_params, log_posterior,
        args=(k_bins, P_obs, P_err, model)
    )

    sampler.run_mcmc(initial_pos, n_steps, progress=True)
    samples = sampler.get_chain(discard=500, thin=10, flat=True)

    return sampler, samples
\end{lstlisting}

\subsection{Simplifications in the Demonstration}
\label{subsec:ps_simplifications}
\begin{itemize}
    \item \textbf{Power Spectrum Model}: Simple parametric form without realistic redshift evolution, scale-dependent bias, or window function effects.

    \item \textbf{Noise Model}: Basic Gaussian noise ignoring correlated residuals from foreground subtraction and instrumental systematics.

    \item \textbf{Parameter Space}: Only 4 parameters versus realistic astrophysical and cosmological parameter spaces with 10+ dimensions.

    \item \textbf{Inference Method}: Basic MCMC without Hamiltonian Monte Carlo, nested sampling, or variational inference optimizations.

    \item \textbf{Data Products}: Single redshift bin analysis without tomographic reconstruction or cross-correlation with other surveys.
\end{itemize}

\subsection{Extensions for Real Array Implementation}
\label{subsec:ps_extensions}
\begin{enumerate}
    \item \textbf{Hierarchical Bayesian Framework}:
    \begin{itemize}
        \item \textbf{Level 1}: Instrumental parameters (gains, beam errors, noise properties)
        \item \textbf{Level 2}: Foreground and systematic error parameters
        \item \textbf{Level 3}: Astrophysical parameters (star formation efficiency, ionizing efficiency)
        \item \textbf{Level 4}: Cosmological parameters ($\sigma_8$, $\Omega_m$, $n_s$, $w$)
    \end{itemize}

    \item \textbf{End-to-end Differentiable Pipeline}:
    \begin{itemize}
        \item Automatic differentiation through entire processing chain (RFI $\to$ calibration $\to$ foregrounds $\to$ power spectrum)
        \item Gradient-based optimization of system parameters against final science metrics
        \item Differentiable cosmological simulations for forward modeling
        \item Uncertainty propagation from raw visibilities to cosmological constraints
    \end{itemize}

    \item \textbf{Cosmological Inference System}:
    \begin{itemize}
        \item Hamiltonian Monte Carlo for high-dimensional parameter spaces
        \item Nested sampling (MultiNest, PolyChord) for evidence calculation
        \item Approximate Bayesian Computation for complex simulation-based likelihoods
        \item Variational inference for real-time parameter constraints
    \end{itemize}

    \item \textbf{Systematic Error Treatment}:
    \begin{itemize}
        \item Nuisance parameter marginalization with priors from instrument characterization
        \item Profile likelihood methods for parameter upper limits
        \item Systematic error templates from end-to-end simulations
        \item Blind analysis techniques to prevent confirmation bias
    \end{itemize}
\end{enumerate}

\section*{Implementation Framework for the Full Array}

\subsection{Computational Infrastructure}
The CosmicWeb-21cm processing pipeline requires a hierarchical computational architecture:

\begin{table}[H]
\centering
\caption{Computational infrastructure requirements}
\label{tab:computational_infra}
\begin{tabular}{@{}lccc@{}}
\toprule
\textbf{Component} & \textbf{Hardware} & \textbf{Latency} & \textbf{Throughput} \\
\midrule
Station Processing & FPGA (Xilinx RFSoC) & $<$10 ms & 100 Gb/s \\
Correlation/Beamforming & GPU Cluster (NVIDIA A100) & $<$1 s & 10 Tb/s \\
Deep Learning Pipeline & HPC (GPU Nodes) & $<$5 min & 1 PFLOPS \\
Cosmological Analysis & HPC (CPU/GPU Hybrid) & $<$1 hr & -- \\
Data Storage & Ceph Object Storage & -- & 100 PB \\
\bottomrule
\end{tabular}
\end{table}

\subsection{Software Architecture}
\begin{itemize}
    \item \textbf{Containerization}: Docker/Kubernetes for reproducible processing environments
    \item \textbf{Workflow Management}: Apache Airflow for pipeline orchestration
    \item \textbf{Version Control}: Git for code and configuration management
    \item \textbf{Continuous Integration}: Automated testing and deployment pipelines
    \item \textbf{Monitoring}: Prometheus/Grafana for system health and performance tracking
\end{itemize}

\subsection{Data Management Strategy}
\begin{itemize}
    \item \textbf{Data Format}: HDF5 with chunked storage for efficient I/O
    \item \textbf{Metadata}: FITS-standard headers with provenance tracking
    \item \textbf{Compression}: Lossless compression (BLOSC) for raw data, lossy compression (JPEG2000) for images
    \item \textbf{Archival}: Tiered storage with tape backup for long-term preservation
    \item \textbf{Distribution}: Data portals with VO-compatible interfaces
\end{itemize}

\subsection{Validation and Commissioning Plan}
\begin{enumerate}
    \item \textbf{Unit Testing}: Individual algorithm validation with synthetic data
    \item \textbf{Integration Testing}: End-to-end pipeline testing with simulated observations
    \item \textbf{Hardware-in-the-loop}: Testing with prototype hardware and real RF environment
    \item \textbf{Cross-validation}: Comparison with independent analysis methods
    \item \textbf{Blind Challenges}: Community-wide analysis competitions
\end{enumerate}

\section*{Conclusion}
These Python demonstrations provide concrete implementations of the core algorithms in the CosmicWeb-21cm signal processing pipeline. While necessarily simplified for clarity and computational efficiency, they capture the essential methodologies that form the foundation of the actual processing system. The extensions outlined for each stage represent the roadmap for scaling these algorithms to handle the full data volume, complexity, and scientific requirements of the CosmicWeb-21cm array.

The co-design philosophy--integrating hardware specifications, observational strategies, and software algorithms--ensures that each processing stage is optimized for the specific challenges of 21cm cosmology. This holistic approach, combined with state-of-the-art machine learning techniques and Bayesian inference methods, positions the CosmicWeb-21cm array to achieve its ambitious scientific goals of probing cosmic reionization, constraining dark energy, and mapping the cosmic web.

The code provided in this appendix serves both as a reference implementation and as a starting point for further development. Researchers interested in adapting these algorithms for other 21cm experiments or related radio astronomy applications are encouraged to use and extend this work, with appropriate attribution.

%------------------------------------------------------------------------------------------------------------------------
% 		ACKNOWLEDGEMENTS
%------------------------------------------------------------------------------------------------------------------------

%\acknowledgements

\end{document}